\documentclass[11pt,a4paper,english]{article}
\usepackage[dvips]{color}
\newcommand{ \beq }{\begin{equation}}
\newcommand{ \eeq }{\end{equation}}

\makeatletter
\@addtoreset{equation}{section}

\makeatother

\begin{document}

\title{\vspace{5cm}{DYNAMICS OF TWO-DIMENSIONAL JOSEPHSON JUNCTION ARRAYS}}
\author{Md. Ashrafuzzaman and Hans Beck\\Institut de Physique\\ Universit\'e de Neuch\^atel, Switzerland\\e-mail: md.ashrafuzzaman@unine.ch\\
\qquad hans.beck@unine.ch}
\maketitle


\section{Introduction}
\subsection{Josephson junction arrays}

Two-dimensional (2D) Josephson junction arrays ($JJA$) offer a unique opportunity for studying a variety of topics in 2D physics, such as phase transitions,
non-linear dynamics, percolation, frustration and disorder, in relatively $\ll$clean$\gg$ experimental realisation. Fabrication of arrays and their basic physical
properties have been described in various articles [1], [2]. A $JJA$ consists of islands, positioned in some periodic or irregular arrangement, that become
superconducting below a given transition temperature $T_c^o$. Below this temperature each island $l$ is characterized by its superconducting wave function
\begin{equation}
\psi=\arrowvert\psi_l\arrowvert e^{i\theta_l}
\end{equation} 
with its amplitude and its phase. For all matters and purposes one can assume that $\arrowvert\psi_l \arrowvert$ has the same value in each island, such that the phase is the only relevant
variable. The islands are linked to each other by the Josephson coupling. The potential energy of the array is then given by 
\begin{equation}
H=\sum_{\langle ll'\rangle}J_{ll'}[1-\cos(\theta_l-\theta_{l'})]
\end{equation}
The sum can usually be restricted to nearest neighbors in the array and the Josephson coupling $J$ is related to the critical current $I_c$ by
\begin{equation}
J=\frac{\hbar}{2e}I_c
\end{equation}
If the array is placed in a perpendicular magnetic field the phase differences in (1.2) have to be replaced by the $\ll$ gauge invariant $\gg$ combination
\begin{equation}
\phi_{ll'}=\theta_l-\theta_{l'}-A_{ll'}
\end{equation} 
where $A_{ll'}$ is the line integral 
\begin{equation}
A_{ll'}=\frac{2e}{\hbar}\int_l^{l'}d{\bf{r}}\cdot {\bf{A}}({\bf{r}})
\end{equation} 
of the applied vector potential. Most of our subsequent calculations will be based on the assumption of a large magnetic penetration depth. Then any vector
potential appearing in our equations describes an applied external electromagnetic field, whereas internal
field fluctuations are neglected. \\

When the phases of neighboring islands differ from each other a supercurrent 
\begin{equation}
K_{ll'}=J_{ll'}\sin\phi_{ll'}
\end{equation}
is flowing through the junction $(l,l')$ and the two islands get charged. The electrostatic energy
\begin{equation}
E_{ll'}=\frac{1}{2}\sum_{ll'}Q_l(C^{-1})_{ll'}Q_{l'}
\end{equation}
has to be added to (1.2). Here $Q_l$ is the charge of island $l$, related to the excess number $\delta N_l$ of Cooper pairs by $Q_l=2e\delta N_l$ , and $C_{ll'}$ is the capacitance matrix of the array.
Its diagonal elements determine the charging energy with respect to ground and the nearest neighbor elements represent the junction capacitance. The excess number
$\delta N_l$ can be identified with a canonical momentum $P_l$ conjugate to the phase [1,3,4]:
\begin{equation}
\lbrack P_l,\theta_{l'}\rbrack=\frac{\hbar}{i}\delta_{ll'}
\end{equation}
The electrostatic energy (1.7) thus plays the role of a (non-local) kinetic energy that has to be added to the potential energy (1.2) yielding thus the complete Hamiltonian
\begin{equation}
H=\frac{1}{2}\sum_{ll'}(2e)^2P_l(C^{-1})_{ll'}P_{l'}+\sum_{\langle ll'\rangle}J_{ll'}(1-\cos\phi_{ll'})
\end{equation}
of the array on which our equations of motion will be based in section 2.\\

Much work has been done in order to elucidate the thermodynamic properties of (1.9) which is indeed a challenging problem of statistical mechanics. The simplest
case taking into account only the potential energy (1.2) in zero applied field is the classical $XY$-model in 2 dimensions. In its ground state all the $\ll$spins$\gg$ are
 parallel, respectively all the phases $\theta_l$ are equal. This system allows to study the critical behaviour of a 2D system with a complex order parameter. It gives rise
to the well-known Berezinski-Kosterlitz-Thouless phase transition [5,6] bringing in the basic concept of vortex excitations. The basic features are summarized in the following subsection.\\

The full form (1.9) adds two complications:\\

a) A $JJA$ in an applied magnetic field corresponds to the $\ll$frustrated$\gg$ $XY$-model. Owing to the additional term $A_{ll'}$ in the coupling energy the phase
configuration in the ground state is more complicated - corresponding to magnetic field induced vortices - and the energy per bond is less negative than $-2J$.
The effect of $A_{ll'}$ is measured in terms of the flux threading through an elementary cell of the array, the $\ll$frustration parameter$\gg$ $f$
\begin{equation}
f=\frac{1}{2\pi}\sum_{\langle ll'\rangle}A_{ll'}
\end{equation}
the sum running over the bonds going around the cell. For simple - rational - values of $f$ the ground state phase pattern is known (see [1] and the references
given there). For certain $f$-values interesting infinite degeneracies have been found in triangular arrays [7] The behavior at finite temperatures, in particular
the various kinds of phase transitions that a frustrated array is supposed to undergo is less well understood [1]. \\

b) Owing to the commutation relations (1.8) the full Hamiltonian (1.9) describes a quantum system, sometimes called $\ll$quantum $XY$-model$\gg$ although its operator
algebra is, of course, different from the $XY$-model describing quantum spins. Quantum fluctuations, surviving even at $T = 0$, add to the thermal fluctuations of the
phases already present in the classical $XY$-model. They tend to counteract the tendency to $\ll$topological$\gg$ phase order which is characteristic of the 2D classical
system. Taking into account only the charging energy with respect to ground the capacitance matrix has the form $C_{ll'} = C \delta_{ll'}$ and the influence of
charging energy can be expressed in terms of the ratio
\begin{equation}
\alpha=\frac{(2e)^2}{2CJ}
\end{equation}

The corresponding phase diagram (with and without dissipation) has been studied by various methods. Approximating the cosine phase interaction by a quadratic form
in the phase differences corresponds to the $\ll$selfconsistent harmonic approximation$\gg$ [4,8,9]. One finds that the critical temperature decreases with
increasing $\alpha$ and finally vanishes at a critical threshold $\alpha_c$ on the order of 6. The same conclusion is drawn from Quantum Monte Carlo calculations [10]. The
quantum model including dissipation is presumably also relevant for describing the superconductor-insulator transition occuring at the underdoped limit of cuprate
superconductors [11]. It has been shown [12] that dissipation may lead to a new universality class characterized by
an effective non-ohmic dissipation.\\

The present article mainly deals with the dynamical properties of $JJAs$. Section 2 presents various equations of motion for the phase variables which are based on
the superconducting properties of the array including dissipation and charging effects. The essential excitations of the phase system are vortices (V) and
antivortices (A) for which we derive appropriate equations of motion in section 3. The following section 4 deals with various linear response functions of an array,
such as its electrical conductance and impedance, and relates such quantities to vortex variables. Various analytical and numerical theoretical techniques that are
currently used in order to calculate dynamic response functions are summarized in section 5. Effects of $\ll$disorder$\gg$, ie deviations from a periodic lattice
structure of the array, are discussed in section 6. Section 7 prepares the interpretation of experimental data by giving an appropriate theoretical description of
the relevant measurements. The main experimental observations of the dynamic behaviour of JJAs are summarized in section 8, whereas section 9  presents
theoretical predictions found by analytical and numerical methods.  Section 10 deals with the dynamic response of $JJAs$ to strong applied currents. The last
section gives some hints to other aspects of $JJA$ that could not be covered in this article. We would like to warn the reader at this point that our presentation has been made to be relatively concise, in order to keep the article at a
decent size. However, we try to give pertinent references where more details can easily be found.\\

\subsection{The Berezinski - Kosterlitz - Thouless scenario}

Treating each phase $\theta_i$ as a classical variable ranging between 0 and $2\pi$ the Hamiltonian (1.2) describes the classical 2D $XY$-model, the phase angle giving the direction
of a classical spin of unit length at site i. Its thermodynamic properties are governed by the Berezinski-Kosterlitz-Thouless $(BKT)$ transition. The lowest energy
configuration corresponds to all phases being equal, corresponding to the $\ll$ferromagnetic$\gg$ ground state of the spins at $T = 0$. For any $T > 0$ this long range order
is replaced by $\ll$topological$\gg$ order due to thermal fluctuations [13]. Besides small amplitude variations, which give rise to $\ll$spin waves$\gg$ in the dynamic
generalization of (1.2), the main finite temperature configurations are given by thermally excited vortices (V) and antivortices (A). Their interaction varies essentially
logarithmically with their distance. They thus behave like a 2D Coulomb gas. In the absence of frustration the latter is neutral (equal number of V and A).
According to the $BKT$ scenario, V and A are bound in pairs below the transition temperature $T_{BKT}$. When $T$ grows above $T_{BKT}$ more and more pairs with large separations
dissociate yielding an increasing number of $\ll$ free $\gg$ V and A. The mean distance between the latter is given by the $BKT$ correlation length $\xi$ with its particular
$T$-dependence
\begin{equation}
\xi(T)=\xi_o~exp(\frac{b}{\sqrt{T/T_{BKT}-1}})
\end{equation}
where $\xi_o$ is a prefactor of the order of the lattice constant and $b$ a parameter of order one. The topological order below $T_{BKT}$ is characterized by a power-law spatial decay of phase (spin)
correlations : 
\begin{equation}
\langle e^{i(\theta_l-\theta_{l'})}\rangle =(\frac{\arrowvert l-l'\arrowvert}{a})^{-\frac{k_BT}{2\pi J(T)}}
\end{equation}
with a coupling constant $J(T)$ renormalized by thermal fluctuations, and a finite helicity modulus $\land$ expressing the $\ll$phase stiffness$\gg$ of the system. $\land$ is given by
the variation of the free energy of the system with respect to a $\ll$phase twist$\gg$ :
\begin{equation}
\land=(\frac{\partial^2F(\alpha)}{\partial\alpha^2})_{\alpha=0}
\end{equation}
where the free energy
\begin{equation}
F(\alpha)=-k_BT \ln\int d[\theta]\quad e^{-\beta H[\theta,\alpha]}
\end{equation}
is evaluated by imposing the boundary condition
\begin{equation}
\theta(la,Na)=\theta(la,0)+\alpha
\end{equation}
relating the phase $\theta(l,m)$ on a given column $l$ of the array at the lower and the upper boundary, respectively, the lattice sites being parametrized in units of the
lattice constant $a$ by $(la,ma)$. $\land$ has a particular temperature dependence. Starting from its $T=0$ value, where it is given by the coupling constant $J$, it is a
deacreasing function of temperature, since thermal fluctuations make the phase system softer. At $T_{BKT}$ it jumps from a universal value
\begin{equation}
\land_c=\frac{2}{\pi}k_BT_{BKT}
\end{equation}
to zero. In section 4 $\land$ will be expressed by suitable phase correlation functions and it will be related to the electrical conductance of the array, as well as to
the vortex dielectric function. This allows to observe the T-dependence of $\land$ by dynamic measurements, as discussed in sections
7 to 9.\\

Above the transition temperature the phase correlation function decays exponentially with the $BKT$ correlation length given by (1.12).\\

In the frustrated case there are magnetic field induced V (or A, depending on the sign of the field) already in the ground state. The thermal V-A pairs coexist with
the latter. The precise nature of the corresponding phase transition in the presence of frustration is still a matter of debate,
 see ch. 8 of ref. [1]. .\\

The precise link between the $XY$-Hamiltonian (1.2) expressed in terms of the phases and the 2-d Coulomb gas can be formalized by transforming it into a vortex
representation with the help of a dual transformation [14, 15].. This procedure allows - at least approximately - to separate the spin wave contribution to the
partition function and yields the vortex Hamiltonian
\begin{equation}
H_c=\frac{1}{2}\sum_{RR'}(m(R)-f)V_c(R-R')(m(R')-f)
\end{equation}

The sum runs over the plaquettes of the array, $m(R)$ being the vorticity of plaquette $R$ (in practice $m = 0$, 1 or -1). The interaction potential $V_c$ depends
logarithmically on the distance, at least for large values of the latter. The frustration $f$ acts on the Coulomb gas of A and V as a $\ll$ background charge $\gg$.
The precise critical behaviour of the $BKT$ transition is obtained by introducing length scale
dependent screening, which yields renormalization group equations [1,13,14] for the relevant parameters of the Coulomb system, namely its dielectric constant,
related to the phase stiffness $\land$ (see section 4) and its fugacity.\\

We do not go into the renormalization group approach to the $BKT$ transition [13] since it concerns mainly thermodynamics. Its generalization to dynamics - our
present topic - as it was developed by Shenoy [16] will be presented briefly in section 5.5. A very extensive review of the $BKT$ transition and the Coulomb gas
model has been published by Minnhagen [17]. Other review articles, such as [18,19], are more directly concerned with dynamic properties of $JJAs$. In order to
interpret correctly measurements on $JJAs$ in the light of the $BKT$ scenario one should, however, bear in mind that the latter (as critical phenomena in general) is in
principle valid for the thermodynamic limit of an infinitely large system. In section 7.1 we will mention finite size effects that can alter the picture to some
extent, in particular through the presence of unbound vortices below the transition temperature.\\

\section{The Dynamics of JJA}

\subsection{Equation of motion for the phases}

In order to derive equations of motion for a $JJA$ we start from our Hamiltonian (1.9) involving Josephson coupling and electrostatic energy. It is useful for later
purposes to include a possible coupling of the array to external currents $I_l$ flowing into (or out of) site $l$
\begin{equation}
H_1=\sum_{l}I_l\theta_l
\end{equation}

The resulting canonical equations read 
\begin{equation}
\dot{\theta}_l =\frac{\partial H}{\partial P_l}=(2e)^2\sum_{l'}(C^{-1})_{ll'}P_{l'}
\end{equation}
\begin{equation}
\dot{P}_l=-\frac{\partial H}{\partial \theta_l}=-\sum_{l'}J_{ll'}\sin(\theta_l-\theta_{l'})-I_l
 \end{equation}

Due to (1.6) and (1.8) the second equation is the continuity relation expressing charge conservation at each site of the array, the right hand side being the sum of
supercurrents and external current flowing into site $l$.\\

Dissipation, due to normal current between islands and from islands to ground, is now easily added to the charge conservation expressed by (2.3) which then reads :
\begin{equation}
\dot{P}_l =-\sum_{l'}J_{ll'}\sin(\theta_l-\theta_{l'})
-\sum_{l'}(R^{-1})_{ll'}(\frac{\hbar}{4\pi e})^2(\dot{\theta}_l-\dot{\theta}_{l'})-R_o^{-1}(\frac{\hbar}{4\pi e})^2\dot{\theta}_l  -I_l
\end{equation}

Here we have introduced the matrix $R_{ll'}$ of junction resistances and the resistance $R_{o}$ to ground, and we have used the Josephson relation between the time derivative
of the phase and a corresponding voltage. Equations (2.2) and (2.4) correspond to a description of the array in terms of an effective circuit involving
capacitive, resistive and Josephson ($\ll$ inductive $\gg$) elements. Neglecting the $\ll$local $\gg$ charging energy and normal current (ie between a given site and the ground)
yields the so-called $\ll$ Resistively and capacitively shunted junction $(RCSJ)$ model $\gg$.\\

Capacitive effects are important for ultrasmall superconducting islands. The interplay between charge and phase fluctuations lowers the superconducting transition
temperature and can drive it completely to zero, leading thus to a quantum superconductor-insulator transition, as discussed in section 1.1. We will mostly
concentrate on dynamic phenomena of $\ll$ classical arrays $\gg$ for which charging effects are unimportant. This corresponds to a motion without inertial acceleration and
thus equation (2.4) with the left hand equal to zero will determine our phase dynamics.  The two most frequently used equations of motion are then distinguished
according to the type of phase damping involved, corresponding to the type of normal current dissipation :   \\

 a. Resistively shunted junction - $RSJ$ - dynamics    \\

Only the $\ll$ bond damping $\gg$ (the second term in (2.4)) is kept, describing dissipation by currents flowing through the junctions. This
dynamics is therefore also called $\ll$ total current conserved $\gg$. With $R_J$ being the junction resistance the $RSJ$ equation for the phases
then reads :
\begin{equation}
\dot{\theta}_l=-\sum_{l'}G_{ll'}(\frac{4\pi e}{\hbar})^2\frac{1}{R_J}\lbrack\sum_{l''}J_{l'l''}\sin(\theta_{l'}-\theta_{l''})+I_{l'}\rbrack
\end{equation}
Here we have introduced the Green function of the lattice which allows to write the sum over bond differences of the phase $\theta$ as a sum
over sites :
 \begin{equation}
\sum_{l'}(\theta_l-\theta_{l'})=\sum_{l'}(G^{-1})_{ll'}\theta_{l'}
\end{equation}

b. Time dependent Landau-Ginzburg - $TDGL$ - dynamics  \\

Taking into account only site damping due to losses to the ground (only $R_o$ is considered in (2.4)) yields the simpler $\ll$relaxational$\gg$
equation of motion
 \begin{equation}
\dot{\theta}_l=-(\frac{4\pi e}{\hbar})^2\frac{1}{R_o}[\sum_{l'}J_{ll'}\sin(\theta_l-\theta_{l'})+I_l]
\end{equation}
This corresponds to the usual time dependent Landau-Ginzburg equation representing $\ll$ nonconserved $\gg$ dynamics (charges are $\ll$ lost $\gg$
by currents going to the ground).\\

In the presence of an electromagnetic field the phase differences occurring in the various current expressions have to be replaced by the corresponding $\ll$ gauge
invariant $\gg$ combinations with the applied vector potential, as indicated in (1.4).\\

Dynamic equations, such as (2.6) or (2.7) can now be used for various analytical considerations, such as the derivation of an equation of motion for the center of a
vortex (section 3) or the calculation of dynamic linear response functions (section 4). They are also at the basis of numerical simulations of equilibrium dynamics
(section 5.4) and of the response of the array to very strong driving currents (section 10).  \\

\subsection{Microscopic derivation of the coupling constants}

The various coupling constants (resistances, capacities and Josephson couplings) of the $RCSJ$ model describe the physical properties of single Josephson junctions.
They can be derived from a microscopic transfer Hamiltonian, describing the coupling between the two superconducting islands through a normal metal or an insulator
by a tunneling interaction and including Coulomb repulsion between electrons. Besides numerous earlier publications on the physics of Josephson junctions useful
reviews, focussing in particular on very small junctions which can now be fabricated, have been given by Sch$\ddot{o}$n and Zaikin [20] and by Simanek [4]. Using either
an imaginary time or a real time representation the quantum mechanical partition function can be rewritten in terms of an action involving the phase difference
between the two superconductors. Its dynamics is indeed governed by a capacitance and by generalized dissipation
and Josephson coupling terms.\\

\section{Vortex dynamics}

\subsection{Classical equations of motion}

Vortex excitations are the main elements determining the physical properties of a $JJA$, in particular in connection with the Berezinski-Kosterlitz-Thouless scenario,
as explained in section 1.2. Thus it is important to account for them explicitly in dynamic calculations. A classical equation of motion for a vortex configuration
can be found, in analogy with the way the dynamics of localized excitations in magnetic models is treated [21]. We start from one of our general equations of motion for the
phases of the array derived in section 2.1 which can be transformed into an equation for the bond supercurrent given by (1.6). In order to prepare a continuum
description of the array it is useful to number each site (of a square array) by an integer $I$ and to attach a horizontal and a vertical bond to each site,
distinguished by the index $s = x$  or $y$, respectively(see figure 1). The bond current $K_{ll'}$ (1.6) is then replaced by $K_s(I)$ and the phase difference (1.4) by $\phi_s(I)$.

\vspace{6cm}

   $Figure~1:$ Parametrization of a square lattice : each site is labelled by an integer $I$. Two nearest neighbor bonds attached
to each site yield the two cartesian directions $x$ and $y$, denotes by the corresponding unit vectors pointing along
the plaquette $I$. They determine the $x$ and $y$ components of vector and tensor quantities in equations (3.1) to (3.5).      \\

Our equations of motion for the phases, derived in section 2, can be used for determining the time evolution of $K_s$
\begin{equation}
C_o\frac{\partial^2K_s(I)}{\partial t^2} =-E_s(I)\sum_{s'I'}\{JM_{ss'}(I-I')K_{s'}(I')+\eta_{ss'}(I-I')\dot{\phi}_s(I')\}
\end{equation}
where $ E_s(I) =\cos\phi_s(I) $ and the $\ll$dynamical matrix$\gg$ $M$ ensures the correct coupling between bonds. \\

In (3.1) we have specialized to the case of a local capacity, but we still admit both types of damping discussed in section $2.1$ :
\begin{equation}
\eta_{ss'}(I-I')=\left\{\begin{array}{ll}
\delta_{ss'}\delta_{II'}\eta_L &\textrm{for local damping}\\

M_{ss'}(I-I')\eta_T & \textrm{for bond damping}
\end{array}\right.
\end{equation}
The friction constants are proportional to the inverse resistances showing up in (2.4), ie the ground (local damping),
respectively the bond resistance (bond damping).\\

A moving vortex is now introduced by a trial solution[22]
\begin{equation}
\phi_s(I,t)=\phi_s^{(v)}(I,t)+\varphi_s(I,t)
\end{equation}
where
\begin{equation}
\phi_s^{(v)}(I,t)=arc\sin K_s^{(v)}(I\arrowvert {\bf{R}}(t))
\end{equation}
The first term represents the circular symmetric $\ll$hedge hog$\gg$ phase pattern with a time dependent center ${\bf{R}}(t)$ and the second contribution describes deviations from
this form. The form (3.2) is substituted into the phase equation of motion (3.1) yielding an equation for the bond currents $K_s^{(V)}$ forming the vortex pattern
\begin{equation}
C_o\frac{\partial^2K_s^{(v)}(I)}{\partial t^2}=-E^s(I)\sum_{s'I'}\{JM_{ss'}(I-I')[E_{s'}(I')\varphi_{s'}(I')+K_{s'}^{(v)}(I')]+\eta_{ss'}(I-I')\frac{\partial}{\partial t}arc\sin K_{s'}^{(v)}(I')\}
\end{equation}
We have neglected the small deformation field $\varphi$ wherever it multiplies terms involving $\phi_s^{(v)}$. The first term on the right hand side of (3.5), involving the dynamical
matrix $M$, represents the Peierls force acting on the moving vortex which originates in the discrete lattice structure of the $JJA$. Indeed, the potential energy of
the phase pattern of the vortex increases when its geometrical center moves away from the middle of a plaquette. It reaches its maximum value when the vortex center
crosses the bond separating two adjacent plaquettes. The height $E_B$ of this energy barrier depends on the lattice structure : $E_B \approx 0.199 J$ for the square
and $E_B\approx 0.043 J$ for the triangular array [23]. Going over from the discrete lattice to a continuum and integrating out the phase variables in a suitable way (see [21]
and [22] for more details) allows to obtain a Newtonian equation for the center ${\bf{R}}$ of the vortex
\begin{equation}
M_v\ddot{\bf{R}}=-\Gamma\dot{\bf{R}}-2\pi qaJ\hat{z}\times{\vec{\varphi}}
\end{equation}
The parameter $q = \pm 1$ determines the sign of the excitation (V or A). The vortex mass $M_v$ is expressed in terms of the grain capacity $C_o$ and $a$ $\ll$ core radius
$\gg$ $b$ of the vortex patterns (which can be taken of the order ot the lattice constant a) :
\begin{equation}
M_v=C_o\frac{20\pi}{9}(\frac{b}{a})^2
\end{equation}
Obviously, in the absence of an underlying discrete lattice structure there is no more energy barrier for the moving vortex. The two types (3.2) of dissipation,
although yielding different equations for the phase field, lead to the same friction term in (3.6), the respective friction constants being only quantitatively different :
 \begin{equation}
\Gamma=\left\{\begin{array}{ll}
\eta_{L}\frac{20\pi}{9}(\frac{b}{a})^2   &\textrm{for local damping}\\

 \eta_{L}\frac{2\pi}{9}(\frac{b}{a})^2 & \textrm{for bond damping}
\end{array}\right.
\end{equation}
One should stress here that different calculations may yield different results for friction constant and mass: $\ll$simple$\gg$ calculations like the above skteched
trial function approach yield a $\ll$bare$\gg$ mass, whereas taking into account the coupling of the vortex configuration with thermal fluctuations introduce a
renormalization to mass and friction, which can even be (logarithmically) divergent due to the 2D $\ll$substrate$\gg$, see, for example, [24]. The second term on the
right hand side of (3.6) can be interpreted in different ways : it can represent the force acting on the vortex when a supercurrent,
given by a phase gradient ${\vec{\varphi}}={\bf{\nabla}}\theta$, is
flowing through the sample. It is usually called "Lorentz force" [1], [25], in analogy to the force a current-carrying wire experiences in a perpendicular magnetic
field. In our later calculations we will insert into equation (3.5) the phase gradient ${\bf{\nabla}}\theta$ resulting from
the presence from another vortex located at ${\bf{R}}_1$ in order
to introduce the effect of V-V interaction in the equation of motion : 
\begin{equation}
M_v\ddot{\bf{R}}=-\Gamma \dot{\bf{R}}+q_1q_2a^2J^2\frac{{\bf{R}}-{\bf{R}}_1}{({\bf{R}}-{\bf{R}}_1)^2}
\end{equation}
When capacitive effects in the array can be neglected (for sufficiently large capacities or for low enough frequencies) the mass
term disappears from (3.6).\\

\subsection{Quantum dynamics}

 At sufficiently low temperatures the $JJA$, characterized by the Hamiltonian (1.9) and an appropriate dissipation function, should be treated by quantum mechanics. As
sketched in section 1.1 this is necessary when the $\ll$mass term$\gg$ in (1.9) becomes comparable to the Josephson coupling, ie when the $\ll$quanticity ratio$\gg$
$\alpha$ given by (1.11) is large enough. \\

The corresponding dynamics of quantum vortices has been studied on the basis of a quantum action for the phases, which allows to treat dissipation on the same
footing as the Hamiltonian terms. Vortex variables are introduced in two ways. Generalizing the Villain transformation, used for the same purpose for the classical
$XY$-model [14] leads to dynamic plaquette vorticities [26]. The transformation is performed on the quantum form of the partition function involving an action
including charging energy and dissipation. As a result one obtains a new action in which the (true) electric charges interact with V and A, acting like
$\ll$topological charges$\gg$. Integrating out the electrical charges yields an effective Lagrangian for the V-A system that attributes effective mass and friction to
the latter. Alternatively, the action is expressed in terms of vortex positions and momenta which are again introduced by a trial form for the dynamic phase field
of a vortex [27], as it was done above for the classical case.               \\

Quantum vortices having a non-zero mass show interesting quantum effect[25], in particular when the lattice pinning potential is
taken into account. While classical particles would oscillate in the minima of this potential and jump across barriers by thermal
excitation, quantum vortices can tunnel through pinning barriers. Moreover, the quantum states of a vortex in the lattice potential will
be a Bloch state with some given momentum. As in the case of band electrons Bloch oscillations can occur, as the particles move across
the Brillouin zone, being driven by en effective electric field. Quantum interference and quantum localization in a random pinning
potential (see also at the end of subsection 5.3) are another manifestation of the quantum nature of vortices. \\

We do not give more details here, since our subsequent analysis will mainly be focussed on $\ll$ classical $\gg$ arrays.     \\

\section{Linear response of a JJA}

\subsection{Kubo formalism for conductance and impedance}

In this section we prepare the theoretical description of experiments that $\ll$ test $\gg$ the dynamic behaviour of a $JJA$. This can be done in different ways by applying
suitable external disturbances to the system. \\

The frequency dependent conductance $G(\omega)$ is obtained from measuring the current that flows in the array as a response to a small time dependent external voltage.
Alternatively, when a weak external current is fed into the array the impedance $Z(\omega)$ determines the resulting voltage drop across it. Explicit expressions for $G(\omega)$
and $Z(\omega)$ in terms of suitable time dependent correlation functions are obtained in this subsection in the framework of linear response theroy. In subsection 4.2 the
response functions $G(\omega)$ and $Z(\omega)$ will be expressed in terms of quantities characterizing the vortex system. Other experimental approaches, such as the two-coil
technique or the measurement of flux noise, will be described in section 7. \\

Since the $JJA$ dynamics involves dissipation it is convenient to interpret the equations of motion (2.3, 2.5) as Langevin equations for the phase and the conjugate
momentum by adding a random force. For calculating response functions it is useful to go over to the corresponding Fokker-Planck $(FP)$ equation [38,39], governing
the probability distribution $\rho(\theta_1,...;P_1,..)$ of the phases and the momenta :
\begin{equation}
 \frac{\partial\rho}{\partial t}=L_{FP}\rho
\end{equation}
The $FP$-operator $L_{FP} = L + L_{ex}$ has a contribution $L$ containing convective and dissipative terms present for an isolated array and $L_{ex}$ describing the coupling to
external sources. Besides the coupling to external currents $I_l$ in (2.1) an external electric field can be applied through a time dependent vector potential $A_{ll'}$ in
the phase differences (1.4) yielding the following form of $L_{ex}$
\begin{equation}
L_{ex}\rho=-\sum_lI_l\frac{\partial\rho}{\partial P_l}+\sum_{ll'}J_{ll'}\cos(\phi_{ll'})\frac{\partial\rho}{\partial P_{l'}} -\sum_{ll'}(\frac{\hbar}{4\pi e})^2R_{ll'}^{-1}\dot{A}_{ll'}\frac{\partial\rho}{\partial P_{l'}}
\end{equation}
Each term in (4.2) has the general structure $L_{ex} = X*F(t)$ where $X$ is an observable of the array expressed in terms of the angles $\theta_l$ and the
conjugate momenta $P_l$ and
$F(t)$ is a time dependent external force $F(t)$. In the $FP$ formalism the linear response of an observable $O$ in the presence of such an
external perturbation is given by splitting the probability distribution
\begin{equation}
\rho=\rho_o+\rho_1
\end{equation}
into the canonical equilibrium part $\rho_o$   and the deviation $\rho_1$ due to $F$. They are, respectively, given by
\begin{equation}
\begin{array}{rcl}
  \rho_o = \frac{1}{Z}  exp(-\beta H) \cr     \\
  \qquad \rho_1(t)=\int_{-\infty}^td\tau~ e^{(t-\tau)L}L_{ex}(\tau)\rho_o
\end{array}
\end{equation}
assuming that the perturbation is adiabatically switched on at $t = -\infty$. The expectation value of $O$ is then given by
\begin{equation}
\langle O(t)\rangle=\int_{-\infty}^td\tau ~R(t-\tau)F(\tau)+O_o
\end{equation}
where $O_o$ is the equilibrium expectation value of $O$. The frequency Fourier transform of the response function $R$ is given by
the retarded time dependent correlation function of the variables $O$ and $X$.
\begin{equation}
 R(\omega)=\int_0^{\infty}dt~ e^{i\omega t}Tr(Oe^{Lt}X\rho_o)
\end{equation}
the symbol $\ll$ Tr $\gg$ meaning integration over phase and momentum variables.\\

We now apply this general scheme to the two experimental configurations mentioned above : \\

a) For calculating the conductance $G(\omega)$ the observable $O$ is the current (supercurrent plus normal current) flowing through
some junction
\begin{equation}
O=J_{ll'}\sin\phi_{ll'}+(R^{-1})_{ll'}(\frac{\hbar}{4\pi e})^2\dot{\phi}_{ll'}
\end{equation}
The action of $L_{ex}$ is given by the second and the third term of (4.2) which, according to (4.4), will act on $\rho_o$.
The derivative with respect to $P_l$ is easily evaluated using  (4.4), (1.9) and (2.2) :
\begin{equation}
\frac{\partial\rho_o}{\partial P_l}=-\beta\rho_o\sum_{l'}(C^{-1})_{ll'}P_{l'}=-\beta\rho_o\dot{\theta}_l
\end{equation}

We specialize to the situation where the applied electric field $E_x$ is homogeneously spread across the $N_x$ sites of the array, say in $x$-direction. Thus the time
derivative ot the vector potential applied to a bond going from site $l$ to site $l'$ in the (positive) $x$-direction is given by
\begin{equation}
\dot{A}_{ll'}=-\frac{E_x}{N_x}=-\dot{A}_{l'l}
\end{equation}

As for calculating the static helicity modulus it is important to impose suitable boundary conditions on the phases for the
conductance calculation. Since we deal with the variables $\theta _1$ and $P_l$ in the present context we use periodic boundary conditions $(PBC)$ for the phases.
Later on, in section 5.4 on numerical simulations, boundary conditions will be introduced which are more appropriate when the relevant variables are vortex positions.
Taking (4.9) and the $PBC$ into account sums like
\begin{equation}
S=\sum_{\langle ll' \rangle_x}(\dot{\theta}_l-\dot{\theta}_{l'})
\end{equation}
extending over all bonds in $x$-direction vanish. Thus the total normal current in (4.7) reduces to
\begin{equation}
     I_{x}^{(n)}=-\eta\sum_{\langle ll' \rangle} \dot{A}_{ll'}=\eta E_x
\end{equation}
which gives the ohmic contribution to the conductance. Moreover, for a regular array the parameters $J_{ll'}$ and $\eta_{ll'}$ are equal for all bonds $\langle ll'\rangle$
so that, due again to (4.9) and the $PBC$ the third term in (4.2), representing the coupling of the applied vector potential in the normal current, also vanishes. We only have to evaluate the response of
the supercurrent in (4.7) to the second term of the external perturbation in (4.2). It has two contributions :
 \begin{equation}
 \langle K_x(t)\rangle=J\sum_ {\langle ll' \rangle_x}\langle\cos\phi_{ll'}\rangle A_{ll'}(t)
 -\beta \frac{J^2}{N_x}\int_{-\infty}^{t}~d\tau  \sum_ {\langle ll' \rangle_x}\sum_ {\langle nn' \rangle_x}\langle\sin\phi_{ll'}(t)\sin\phi_{ll'}(\tau)\dot{\phi}_{nn'}(\tau)\rangle A_{nn'}(\tau)
 \end{equation}

Partial integration of the second term and Fourier transforming with respect to time leads to the final result  
\begin{equation}
  \langle K_x(\omega)\rangle=  G_s(\omega)E_x
\end{equation}
where the superconducting part of the array conductance
\begin{equation}
G_s(\omega)=\frac{\land}{i\omega}-\beta F_c(\omega)
\end{equation}
is expressed in terms of the helicity modulus
 \begin{equation}
  \land=J\sum_ {\langle ll' \rangle_x}\langle\cos\phi_{ll'}\rangle-\beta\frac{J^2}{N_x} \sum_ {\langle ll' \rangle_x}\sum_ {\langle nn' \rangle_x} \langle\sin\phi_{ll'}\sin\phi_{nn'}\rangle
 \end{equation}
and a dynamic current-current correlation function
\begin{equation}
F_c(\omega)=\int_{0}^{\infty}dt~e^{i\omega t}J^2\sum_{\langle ll'\rangle_x}\sum_{\langle nn'\rangle _x} \langle\sin\phi_{ll'}(t)\sin\phi_{nn'}(0)\rangle
\end{equation}
Finally, the total conductance $G(\omega)$ is given by
\begin{equation}
   G(\omega)=\frac{\land}{i\omega}-\beta F_c(\omega)+\eta
\end{equation}

The superconducting contribution to $G$ has already been derived by Shenoy [28].

Thus the dynamic conductance $G$ is expressed in terms of an effective circuit with parallel superconducting and normal channels. This picture can be made more
explicit by writing $G$ as
\begin{equation}
   G(\omega)=\frac{1}{i\omega L(\omega)}+\frac{1}{R(\omega)}
\end{equation}
with two real, frequency dependent response functions : an effective inductance L and an effective resistance $R$. Comparison of (4.17) with (4.18) shows that the
helicity modulus which is one of the basic features of the $BKT$ transition can be extracted from the low frequency behaviour
 of the dynamic conductance.\\

 b) The impedance $Z(\omega)$ is obtained by calculating the expectation value of the voltage
\begin{equation}
 V_{ll'}=\frac{\hbar}{2e}(\dot{\theta}_l-\dot{\theta}_{l'})
\end{equation}
across a given junction under the influence of an applied current acting through the first term of $L_{ex}$ in (4.2). In order
to allow the external current to enter the
array open boundary conditions have to be used for this calculation. Again using (4.8) for the derivative of $\rho_o$ in (4.3),
the response function is immediately found to be
\begin{equation}
 \langle V_{ll'}(t)\rangle=\int_{-\infty}^t d\tau\sum_n\langle(\dot{\theta}_l(t)-\theta_{l'}(t))\dot{\theta}_n(\tau)\rangle I_n(\tau)
\end{equation}

The full impedance is found by summing the voltage drops over all bonds across the sample in $x$-direction and letting an equal current $I$ flowing in on the left side
through the sites $(0, ma)$, respectively out on the right sides $(N_xa, ma)$ of the array. This yields the final expression for the impedance
\begin{equation}
Z(\omega)=\frac{\hbar}{2e}\sum_{\langle ll' \rangle _x}\int_0^{\infty}dt~e^{i\omega t}\langle(\dot{\theta}_l(t)-\dot{\theta}_{l'}(t))
(\dot{\theta}_{0,ma}-\dot{\theta}_{N_xa,ma})\rangle
\end{equation}

The expressions for $G$ and $ Z $ in terms of the corresponding dynamic correlation functions do not seem to have much in common. It is thus legitimate to ask the
question whether the expected relation 
\begin{equation}
 G(\omega)=Z(\omega)^{-1}
\end{equation}
holds. Expressing these correlation functions in the following subsection, for a classical array, by suitable vortex variables will indeed confirm (4.22).
Alternatively, expressing the partition function of an array including capacitive effects and damping terms as an imaginary time functional integral, and going to a
dual representation using the electrical charge on each site rather than the phase as variables, Panykov and Zaikin [29] have shown that the corresponding linear
response expressions for $G$ and $Z$ indeed satisfy (4.22). However, relation (4.22), as plausible it may look, need not necessarily hold. Indeed, it has been
shown [29,30] that for a quantum array the linear voltage response to an external current is not equivalent to the inverse of the linear current
response to an applied voltage. The difference between the two arises because the voltage bias suppresses some important quantum fluctuations of the phase
(instantons) due to the fact that the total phase difference across the sample is fixed by the applied voltage drop. Therefore the phases are hindered in their
quantum fluctuations. The latter are not suppressed by an applied current since open boundary conditions are then used.
As a result, the impedance $Z$ sees the sample
$\ll$ less superconducting $\gg$ than the conductance ! The same effect could, in principle, also occur with thermal fluctuations
in the classical regime, but an explicit
example seems not to have been constructed. \\

\subsection{Dynamic correlation functions and dielectric function for the vortex system}

As stated in section 1.2 our vortex-antivortex system is modelled by the 2D neutral Coulomb gas (2DCG). It is therefore useful to introduce some concepts that are
currently used when treating the dynamics of the 2DCG (see, for example, the chapter $\ll$Charged fluids$\gg$ of ref. [31]). For a hydrodynamic description, adapted to
treat various $\ll$macroscopic$\gg$ experiments treated in section 7, the basic variables are charge density $\rho({\bf{r}})$ and current
density ${\bf{j}}({\bf{r}})$, the spatial Fourier transforms
of which are related to the positions ${\bf{R}}_l$ and the sign $q_l = \pm 1$ of the charge of the $N$ particles by
\begin{equation}
\begin{array}{rcl}
 \rho({\bf{p}})=\frac{1}{\sqrt{N}}\sum_lq_le^{i{\bf{p}}\cdot{\bf{R}}_l}\cr\\
{\bf{j}}({\bf{p}})=\frac{1}{\sqrt{N}}\sum_lq_l\dot{\bf{R}}_le^{i{\bf{p}}\cdot{\bf{R}}_l}
\end{array}
\end{equation}

In section 5.3 we will use Mori's formalism in order to calculate dynamic correlation functions of these basic observables, such as the $\ll$charge density correlator$\gg$
\begin{equation}
   \phi_{\rho\rho}({\bf{p}},z)=\int_0^{\infty}dt~e^{-zt}\langle\rho({\bf{p}},t)\rho^*({\bf{p}},0) \rangle
\end{equation}
where the average is over a canonical ensemble and the time dependent particle positions have to be obtained by integrating suitable equations of motion. The equal
time correlation function 
\begin{equation}
 S_c({\bf{p}})=\langle\arrowvert\rho({\bf{p}})\arrowvert^2\rangle
\end{equation}
is the charge structure factor of the 2DCG. $\phi _{\rho\rho}$ and $S_c$ are related to the dynamic charge susceptibility $\chi$ by [31]
\begin{equation}
  \chi({\bf{p}},z)=-\frac{1}{k_BT}[S_c({\bf{p}})-z\phi_{\rho\rho}({\bf{p}},z)]
\end{equation}
which, in turn, is linked with the dynamic dielectric function $\epsilon(\omega)$ by
\begin{equation}
 \frac{1}{\epsilon(\omega)}=\lim_{p\to 0}[1+\frac{2\pi q_o^2n}{p^2}\chi({\bf{p}},-i\omega)]
\end{equation}
where $q_o$ is the charge of the particles and $n$ their density. As usual in electrodynamics the dielectric function can be
expressed in terms of a dynamic conductance $\sigma_v(\omega)$:
\begin{equation}
  \epsilon(\omega)=1-2\pi\frac{\sigma_v(\omega)}{i\omega}
\end{equation}

Here we have used the letter $\sigma$ and a subscript $v$ in order to distinguish this $\ll$ vortex conductance $\gg$, related to the  mobility of the vortices and thus to their response to
an applied force, from the physical electric conductance $G$ of the $JJA$.  \\

For many purposes - in particular in order to illustrate the relation $G(\omega)=Z(\omega)^{-1} $- it is now useful to express
our response functions directly by these dynamic
vortex correlation functions. Various procedures have been proposed in order to transform the $XY$ problem from its phase form
(expression (1.2)) into a vortex representation. An approximate dual transformation allows to separate linear (spin wave) and vortex excitations, yielding the vortex
Hamiltonian (1.18). This is particularly useful for dealing with equilibrium thermodynamics, but the same approach has also been applied to the action in a functional
integral representation of the quantum version of the $XY$-model.     \\

Another approach yielding insight in the dynamics of the system [33] consists in approximating the lattice phase Hamiltonian by a an energy functional for
continuous phase and momentum fields 
\begin{equation}
 H=\frac{J}{2}\int d^2r(\nabla\theta)^2+\frac{1}{2C_o}\int d^2r{\bf{P}}(r)
\end{equation}
and separating the phase field $\theta$ into a $\ll$ regular $\gg$ (spin wave) part $\phi$ and a vortex part $\psi$, given by
\begin{equation}
 \theta=\phi+\psi;~\nabla\times\nabla\phi=0,~\nabla\cdot\nabla\psi =0
\end{equation}

C$\hat{o}$t$\acute{e}$ and Griffin [33] transform the equations of motion for the phase and the conjugate momentum fields, without dissipative terms, into Maxwell equations in which
the vortices act like effective charges being the source of effective electric and magnetic fields related to the phase field by 
\begin{equation}
\begin{array}{rcl}
{\bf{E}}= \sqrt{2\pi }\nabla\theta\times\hat{z}\cr  \\
{\bf{B}}=\sqrt{\frac{2\pi}{C_o}}p\hat{z}
\end{array}
\end{equation}
$\hat{z}$ being a unit vector perpendicular to the array. The charge of a vortex is related to the original coupling constant in (4.29) by
\begin{equation}
   q_o^2=2\pi J
\end{equation}
This concept can easily be generalized to include dissipation by adding a local damping term (normal currents flowing to the ground) to the equation of motion as in
section 2. Faraday's equation is then generalized to
\begin{equation}
\frac{1}{c}\dot{\bf{B}}=-\nabla\times{\bf{E}}-\frac{1}{cC_oR_o}{\bf{B}}
\end{equation}
with
\begin{equation}
c=\sqrt{\frac{J}{C_o}}
\end{equation}
being the $\ll$ speed of light $\gg$ in this effective electrodynamics of the array. Expressing super and normal current densities by
\begin{equation}
\begin{array}{rcl}
{\bf{K}}_s=J\nabla\theta \cr  \\
{\bf{I}}_n=\frac{\hbar}{2e}\frac{1}{R_o}\nabla\dot{\theta}\equiv\eta\nabla\dot{\theta}
\end{array}
\end{equation}
allows now to express the response function found above for $G$ and $Z$   in terms of vortex quantities.                        \\

For the impedance $Z$ we have to establish a linear relation between the total current, ${\bf{I}} = {\bf{K}}_s +{\bf{I}}_n$, and the voltage across the sample. We start from Maxwell's
equation 
\begin{equation}
   \nabla\times{\bf{B}}=\frac{1}{c}\dot{\bf{E}}-\frac{2\pi}{c}{\bf{j}}_v
\end{equation}
${\bf{j}}_v$ being the vortex current density and rewrite it in terms of the time derivative of the phase gradient
\begin{equation}
\nabla\dot{\theta}=\frac{1}{\sqrt{2\pi cC_o}}(\hat{z}\times\dot{\bf{E}}-2\pi\hat{z}\times{\bf{j}}_v)
\end{equation}
Using (4.31) and (4.35) we replace the effective electric field by $K_s = I - K_n$. Then we relate, within our $\ll$effective electrodynamics$\gg$, the vortex current
density ${\bf{j}}_v$ to the driving electric field through the vortex conductance $\sigma_v$ :
\begin{equation}
 {\bf{j}}_v= \sigma_v{\bf{E}}
\end{equation}
which yields
\begin{equation}
  (1+i\omega\frac{\eta}{J}-\eta\frac{2\pi\sigma_v}{J})\nabla\dot{\theta}={\bf{I}}(\frac{i\omega}{J}-\frac{2\pi\sigma_v}{J})
\end{equation}

The dynamic vortex dielectric function is introduced by (4.28). Using the dynamic Josephson relation
\begin{equation}
 V=\frac{\hbar}{2e}\nabla\dot{\theta}
\end{equation}
we finally express the impedance $Z = \frac{V}{I}$ in terms of $\epsilon_v(\omega)$:
\begin{equation}
    Z(\omega)=\frac{\hbar}{2e}\frac{1}{\eta+\frac{J}{i\omega\epsilon_v(\omega)}}
\end{equation}

For the dynamic conductance, given by (4.17), we first note that the helicity modulus $\land$ is directly related to the static vortex dielectric function[15]:
\begin{equation}
         \land=\frac{J}{\epsilon_v(\omega=0)}
\end{equation}

In the continuum description the supercurrent density is given by (4.35). For our $PBC$ the regular part of the phase gradient does not contribute to the dynamic
current-current correlator in (4.16). According to (4.31) the singular part stands for the effective electric field related to the vortex charge density by
Poisson's equation. Using also $q_o^2=2\pi J$ this allows to relate (4.16) to the dynamic vortex charge correlator (4.24) :
\begin{equation}
  F_c(\omega)=\beta J^2(2\pi)^2n_v\lim_{p\to 0}\frac{\phi_{\rho\rho}({\bf{p}},-i\omega)}{p^2}
\end{equation}
$n_v$ being the vortex number density. The way towards the vortex dielectric function is found by using the relations (4.26) and (4.27), once for finite
frequency $\omega$ and once for $\omega=0$, which immediately yields
\begin{equation}
  F_c(\omega)=\frac{J}{i\omega}[\frac{1}{\epsilon_v(\omega)}-\frac{1}{\epsilon_v(0)}]
\end{equation}

Substituting this result and (4.17) in the expression (4.20) for $G$ leads to
\begin{equation}
 G(\omega)=\frac{J}{i\omega\epsilon_v(\omega)}+\eta
\end{equation}
Manifestly it is the reciprocal of the impedance given by (4.41). Expression (4.45) is a particular case of the form (4.18), in which the original bond resistance
becomes frequency dependent through the vortex dielectric function. Expressing the latter by the vortex conductivity through (4.45) one obtains
 \begin{equation}
 G(\omega)=\frac{J}{2\pi\sigma_v(\omega)+i\omega}+\eta
\end{equation}

Thus, the superconducting contribution to $G$ is directly related to the vortex mobility : $G(0)$ is inversely proportional to the latter. In particular, when the
vortex system has a non vanishing mobility $G(0)$ is finite and superconductivity is destroyed.

\section{Theoretical techniques for calculating dynamic vortex response function}

\subsection{Early approaches}

Ambegaokar et al[34] calculated the dielectric function $\epsilon(\omega)$ of the vortex system
 \begin{equation}
   \epsilon(\omega)=1+2\pi\chi(\omega)
 \end{equation}
for independent bound pairs by averaging the dynamic susceptibility $\chi$ of a dipole of length $r$ over a Boltzmann
probability distribution $\varphi(r,T) = e^{-\beta V(T,r)}$ of
finding a V-A-pair of length $r$ at a given temperature. A simple form for the effective interaction $V$ (the Coulomb interaction devided by a dielectric constant)
yields 
\begin{equation}
    \varphi(r)\propto(\frac{r}{a})^{-4\frac{T_{BKT}}{T}}
\end{equation}

Alternatively one can use a length dependent static dielectric function [34 and 35]. The resulting dielectric function
\begin{equation}
 \epsilon(\omega)=1+\frac{q_o^2}{4a}\int_a^{\infty}rdr\frac{\varphi(r)}{-i\omega\Gamma+\frac{k_BT}{r^2}}
\end{equation}
has a power law dependence on frequency with exponents that depend on temperature.       \\

For the free vortex system, above $T_{BKT}$, the simplest form for $\epsilon$ is given by Drude's form obtained by inserting a frequency independent vortex conductivity $\sigma_o$ which is
proportional to the inverse of the friction function in the equation of motion (3.6). This yields the well-known simple form for the inverse of the dielectric function :
\begin{equation}
\begin{array}{rcl}
Re(\frac{1}{\epsilon(\omega)})=\frac{\omega^2}{\sigma_o^2+\omega^2} \cr    \\
Im(\frac{1}{\epsilon(\omega)})=\frac{\omega\sigma_o}{\sigma_o^2+\omega^2}
\end{array}
\end{equation}

\subsection{Minnhagen's phenomenology (MP)}

This approach [17] starts from a Debye form of the wave number dependent static dielectric function
$\frac{1}{\epsilon(k)} =\frac{k^2}{k^2 + k_s^2}$ with a screening wave vector $k_s$ and goes to $Re(1/\epsilon(\omega))$ by replacing
the wave number $k$ by frequency $\omega$ through
\begin{equation}
  k^2=\frac{\omega}{D}
\end{equation}
which corresponds to taking into account diffusive motion of the particles with diffusivity $D$. Performing a Kramers-Kronig transformation one obtains
 \begin{equation}
 \begin{array}{rcl}
    Re(\frac{1}{\epsilon(\omega)})=\frac{\omega}{\omega+k_s^2D}\cr      \\
Im(\frac{1}{\epsilon(\omega)})=\frac{2}{\pi}\frac{\omega Dk_s^2}{\omega^2-(Dk_s^2)^2}\ln\frac{Dk_s^2}{\omega}
 \end{array}
 \end{equation}
Since $\epsilon(\omega) = 1 + \sigma(\omega)/(-i\omega)$ this amounts to using a vortex conductivity with a peculiar non-analytic frequency dependence for
small $\omega$
 \begin{equation}
 \begin{array}{rcl}
    Re\sigma(\omega)=-\frac{\pi Dk_s^2}{2}(\ln\frac{Dk_s^2}{\omega})^{-1}\cr        \\
    Im\sigma(\omega)=-\frac{\pi ^2Dk_s^2}{4}(\ln\frac{Dk_s^2}{\omega})^{-2}
 \end{array}
 \end{equation}

At first sight one would tend to reject this result, since - had one made the more correct replacement 
 \begin{displaymath}
           k^2=\frac{i\omega}{D}
 \end{displaymath}
as one should for a diffusive dynamics - one would come back to Drude's form rather than to (5.6) and (5.7). It turns out, however, that simulations and
experimental data confirm the $MP$ form of the dynamic response in an appropriate frequency domain, as we will show in sections 8 and 9.\\

A simple criterion allows to distinguish the $MP$ form (5.6) from the more simple Drude expression : the $\ll$peak ratio$\gg$
 \begin{equation}
 r=\frac{Im(\frac{1}{\epsilon(\omega_o)})}{Re(\frac{1}{\epsilon(\omega_o)})}
  \end{equation}
between imaginary and real parts of the inverse dielectric function, evaluated at the frequency $\omega_o$ where $Im(1/\epsilon)$ has its maximum, is given by $r = 1$ for Drude's form
and by $r = 2/\pi$ for $MP$ [17].

\subsection{Equations of motion (Mori, Fokker-Planck)}

The dynamic charge density correlator $\phi_{\rho\rho}({\bf{p}},z)$ defined in (4.24) is a central quantity for treating vortex dynamics. Indeed, it is directly related to
experimentally accessible quantities such as the dynamic dielectric function (see section 4.2) and the flux noise spectrum that will be introduced in section 7.3. It
is therefore important to rely on systematic methods for calculating such classical time dependent correlation functions.\\

Mori's approach [36] for evaluating dynamic correlation functions of systems for which relevant static correlation functions are known is particularly useful for
our 2D Coulomb gas. It can be based on the two hydrodynamic variables $\rho$ and ${\bf{j}}$ introduced in section 4.2. The time evolution of these variables is easily found by
using the equations of motion for the center of a vortex (see section 3) including force and friction terms. Mori's technique yields a formally exact expression for
$\phi_{\rho\rho}({\bf{p}},z)$
\begin{equation}
    \phi_{\rho\rho}({\bf{p}},z) =\frac{S_c(p)}{z+\frac{k_BTp^2}{S_c(p)\gamma(p,z)}}
\end{equation}
where $S_c(p)$ is the charge structure factor (4.25) of the vortex system and $\gamma(p,z)$ - the $\ll$ memory function $\gg$ given by the dynamic correlator of the $\ll$random
force$\gg$ - summarizes the complicated effects of the Coulomb interaction on charge dynamics [35]. Various approximations can be used in order to obtain manageable
expressions for $\gamma(p,z)$. We first focus on the region above the $BKT$ temperature treating the vortex system in its $\ll$metallic$\gg$ phase, ie by neglecting possible
binding in pairs. We assume that the zero wave vector limit $\gamma({\bf{0}},z)=\gamma(z)$ is sufficient to describe hydrodynamic phenomena. The general form of this quantity is
given by [35] :
\begin{equation}
  \gamma(z)\approx\Gamma+\frac{q_o^2}{2k_BTn_v\Omega^3}\sum_{{\bf{k}}{\bf{k'}}}{\bf{k}}\cdot{\bf{k'}}V_c({\bf{k}}) V_c^*({\bf{k'}})\int_0^{\infty}dt~e^{-zt}\langle n^*({\bf{k}},t)\rho({\bf{k}},t)n({\bf{k'}},0)\rho({\bf{k'}},0)\rangle
\end{equation}

 Here $\Omega$ is the total area of the system, $V_c(p) =\frac{2\pi q_o^2}{p^2}$ is the Fourier transform of the Coulomb potential and $n({\bf{p}})$ is the particle density
\begin{displaymath}
n({\bf{p}})=\frac{1}{\sqrt{N}}\sum_le^{i{\bf{p}}\cdot{\bf{R}}_l}
 \end{displaymath}
corresponding to the charge density in (4.8). In the above statement we have - as it is usually done - neglected the action of the $\ll$Mori projector$\gg$ [36] in
the time evolution of the higher order correlation function on the right hand side of (5.10). In the absence of interparticle interaction $\gamma(z)$ simply reduces to
the bare friction $\Gamma$. Thus $\gamma(z)$ is a generalized friction function and its inverse is proportional to the $\ll$charge mobility$\gg$ of the system. This can be seen by
inserting the specific form (5.8) of $\phi_{\rho\rho}({\bf{p}},z)$  into the relations (4.26) and (4.27) determining the dielectric function. One finds the form (4.28) with
\begin{equation}
\sigma_v(\omega)=\frac{2\pi q_o^2n_v}{\gamma(-i\omega)}
\end{equation}

This expressions illustrates the role of the inverse of the friction function as a $\ll$charge mobility$\gg$.\\

As usual in treating interacting systems a higher order correlator, involving both charge and number densities, shows up in (5.10). In the spirit of the $\ll$ mode
coupling $\gg$ approximation, currently used in liquid dynamics [37], this higher order correlator is factorized into a product of  $\phi_{\rho\rho}({\bf{p}},z)$ and of the corresponding
number density correlator  
\begin{equation}
  \phi_{nn}({\bf{p}},z)=\int_0^{\infty}dt~e^{-zt}\langle n({\bf{p}}t)n^*({\bf{p}}0)\rangle
\end{equation}

However, in order that such a factorization be a useful approximation, the long range Coulomb potential - which induces long range correlations between particles -
should be replaced by the screened interaction, as it is customary in treating the dynamic properties of Coulomb systems [40,41]
\begin{equation}
 V_{sc}(p)=\frac{2\pi q_o^2}{p^2+p_f^2}
\end{equation}
which, in this simple form, involves the Debye screening wave number
\begin{equation}
p_f^2=\frac{k_BT}{2\pi q_o^2n_f}
\end{equation}
determined by the charge $q_o$ and the density $n_f$ of free vortices. For the static structure factor $S_c(p)$ we use the simple form
 \begin{equation}
   S_c(p)=\frac{p^2}{p^2+p_o^2}
 \end{equation}
with
\begin{displaymath}
 p_o^2=\frac{k_BT}{2\pi q_o^2n_v}
\end{displaymath}
which satisfies the sum rule [31] for the metallic phase of a Coulomb gas. Note, that $p_o$, the relevant wave number for $S_c(p)$, involves the total particle density.
For the number density correlator (5.11) we use a simple diffusion form
\begin{equation}
         \phi_{nn}(p,z)=\frac{n_v}{z+\frac{k_BTp^2}{\Gamma}}
\end{equation}

Equation (5.9) is still an implicit expression for $\gamma(z)$ since the right hand side involves the same quantity again through $\phi_{\rho\rho}({\bf{p}},z)$. To lowest order in the
interaction $\gamma(z)$ can be evaluated by replacing it by $\Gamma$ under the integral on the right hand side of (5.9). Results for $\gamma(z)$ and thus for the flux noise spectrum
and the dielectric function, obtained by using this procedure will be mentioned in section 9.  \\

 Mori's approach also allows to take into account pinning (Peierls) forces acting on the vortices, as they result from equation (3.5)
(see the discussion following that equation). One then obtains another contribution to the friction function $\gamma(z)$, having a similar
structure as (5.10), with the Coulomb force being replaced by the pinning force. This approach has been applied [42] to take into
account the effect of random pinning in disordered lattices (described in section 6.). A mean-field like description of disorder
induced localization of vortices is found in this way [42]. \\

The Fokker-Planck equation, presented in section 4.1 in connection with linear response calculation using phases as variables, can also be used for determining the
vortex distribution function, providing thus another systematic approach for evaluating dynamic correlation functions. It has been used by Timm [43] for
calculating the flux noise spectrum in an independent pair approximation. The corresponding Fokker-Planck operator takes into account the free diffusion of pairs,
the internal interaction between the partners and a possible annihilation of the V and A constituting the pair. The results of this approach will be mentioned in
section 9.

\subsection{Numerical simulations}

The dynamic properties of JJAs have been studied by solving numerically different types of equations of motion:  \\

a) for the phases of the XY-model using RSJ dynamics (see our equation (2.5)) : References 44-54.\\

b) for the phases of the XY-model using TDGL dynamics (see our equation (2.7)) : References 44-46, 50, 53, 55-60.\\

c) for the positions of the charges in the 2D V-A-Coulomg gas (see our equation (3.9)) : References 59, 61-63.\\

d) for the 2D lattice Coulomb gas using Monte Carlo dynamics : References 64, 65.    \\

Details about the simulation technique can be found in these references. The correct choice of boundary conditions is particularly important. Indeed, various
quantities that are relevant for the $BKT$ transition, such as the phase stiffness, describe the response of the system to forces applied to its boundaries. For
simulations of the $XY$-model (a and b) periodic boundary conditions $(PBC)$ imposed in the phase angle $\theta_l$ are most commonly used when the total current in the sample
is zero (see our discussion of boundary conditions in the context of linear response calculations in section 4.1). However, as shown by Olsson [66], $PBC$ for the
phase angles lead to non-periodic boundary conditions for the vortex interaction. On the other hand, Coulomb gas simulations (c and d) prefer $PBC$ for the motion of the point
particles representing the vortices. This requires imposing $\ll$fluctuating twist boundary conditions$\gg$ $(FTBC)$ on the phase angles [66]. In this case a variable $D =
(D_x,D_y)$ is added to the phase difference (1.4) between neighboring sites $l$ and $l'$ :
\begin{equation}
\phi_{ll'}\to\phi_{ll'}-{\bf{e}}_{ll'}\cdot{\bf{D}}
\end{equation}
the vector ${\bf{D}}$ being projected onto the unit vector ${\bf{e}}$ of the bond ($ll'$). For the phase angles themselves $PBC$ are used. ${\bf{D}}(t)$ is itself a dynamic quantity which is
related to vortex motion : whenever a vortex has crossed the sample of length $L$ from one side to the other ${\bf{D}}$ changes by $2\pi/L$. Its equation of motion is established
by considering [45] current conservation across the sample 
\begin{equation}
   \frac{V_x}{R} +J\sum_{\langle ll'\rangle _x}\sin(\phi_{ll'}-D_x)+\eta_x=0
\end{equation}

Here, the first term expresses the normal current in $x$-direction in terms of the voltage drop $V_x$ and the normal resistance $R$, and the last term represents the
thermal noise current used in the usual Langevin equation in order to impose a given temperature $T$ on the system. By the dynamic Josephson relation the voltage is
proportional to the time derivative of the total phase change across the sample. Since $PBC$ are imposed on $\theta_l$ it is directly related to the time derivative of
$D_x$ :
\begin{equation}
 V_x=-\frac{\hbar}{2e}L\dot{D}_x
\end{equation}

Combining these relations leads to an equation of motion for ${\bf{D}}$ [66]. By considering the Fokker-Planck equation for the probability distribution of the phase
angles and the additional variables $(D_x,D_y)$ one obtains the standard white noise correlations for the noise current :
\begin{equation}
   \langle\eta_x(t)\eta_x(0)\rangle=\frac{2T}{L^2}\delta(t)
\end{equation}

A current passing through the sample can either be incorporated in the current conservation condition that determines ${\bf{D}}(t)$, or - alternatively - open boundary
conditions for the phase angles can be used (see section 4.1). The former approach has the advantage that one can continue to use periodic boundary conditions, which
is more convenient for numerical calculations.

\subsection{Dynamic scaling}

The $BKT$ scenario governs the phase transition in a 2D model involving a complex order parameter with an amplitude and a phase. It has the particularity that the
correlation length, diverging as usual when $T$ goes to its critical value from above, stays infinite for all $T<T_{KT}$. Thus the usual dynamic scaling relation
 \begin{equation}
 \begin{array}{rcl}
      F(k,\omega)=\hat{F}(k\xi,\omega\xi^z)  \cr      \\
      F(r,t)=\hat{F}(r/\xi,t\xi^{-z})
 \end{array}
 \end{equation}
for functions depending on space and time variables, is valid a priori only above $T_{BKT}$. The dynamic critical exponent $z$ summarizes the essential information about
vortex dynamics. Numerous calculations, mainly numerical, have been aimed at obtaining $z$. Apriori, one expects $z = 2$ [50], according to the usual classification
of dynamic critical phenomena [67], which is indeed bourne out by various calculations, and it is also reflected by the flux noise measurements of ref [68]. It
corresponds to the fact that motion of vortices is of diffusive nature.   \\

Below $T_{BKT}$, where the model remains critical, the situation is more complex. Since $ \xi=\infty$ the relevant scaling now involves other typical lengths, such as the
sample size or the size of the detection device (such as the coil used for flux noise measurement, see section 7). Morevoer, $z$ depends on the dynamic phenomenon to
which it is applied and thus on the boundray conditions used. Jenssen et al [50] give a rather complete overview over previous results and over their own
calculations. They have used $RSJ$, as well as $LG$ dynamics for the phases (see section 2), applying, for both cases, either periodic or fluctuating twist boundary
conditions and they have analysed resistance, supercurrent scaling and the short time relaxation of a non-equilibrium configuration. \\

An interesting observation has recently been made for the three dimensional(3D) $XY$-model : there is evidence [59] for the existence of two dynamic critical exponents signalling the dependence of
dynamic correlation functions on wave number $k$ and on the correlation length $\xi$:
\begin{equation}
    F(k,t)=\hat{F}(k\xi,t\xi^{-z_o},tk^{-z})
\end{equation}
with $z_o=1.5$ and $z=1$, respectively $z=2$ for the 3D lattice Coulomb gas, respectively the 3D $XY$-model. \\

From an experimental point of view a particularly interesting application relates the value of $z$ below $T_{BKT}$ to the non-linear current-voltage characteristics
$V\sim I^{a(T)}$ (see section 7.1). Ambegaokar et al [34] found
\begin{equation}
   a(T)=\frac{\pi J(T)}{k_BT}+1
\end{equation}
where $J(T) = J/\epsilon(T)$ is the effective $XY$-coupling constant, renormalized by spin wave and VA-pair fluctuations represented by the static
dielectric constant $\epsilon(T)$. The non-linear I-V relation relies on the fact that VA pairs that are separated by an applied current (and thus contribute to a finite voltage like $\ll$ free $\gg$
vortices) can recombine, see section 7.1. Expression (5.22) is found be setting the pair breaking and the recombination rates equal to each other [34].
At the critical temperature one obtains $a = 3$, which together with $z = 2$ satisfies the dynamical scaling
relation [69] $a = z + 1$. More recently Minnhagen et al [49] have found a different value
 \begin{equation}
         \hat{a}(T)=\frac{2\pi J(T)}{k_BT}-1
\end{equation}
by combining static and dynamic scaling arguments. They then link this expression to a temperature dependent dynamical exponent
\begin{equation}
   \hat{z}(T)=\hat{a}(T)-1
\end{equation}
Interestingly enough, both expressions yield the same value $a=3$ at $T_{BKT}$, since (see section 1) :
\begin{equation}
      J(T_{BKT})=\frac{2k_BT_{BKT}}{\pi}
 \end{equation}

Bormann [70] has established a link between (5.23) and a more involved recombination scenario - a collective $\ll$partner transfer$\gg$ between bound pairs - which is
supposed to take place when the vortex density is not too small. A recent analysis of I-V data obtained for various 2D superconductors [71] has produced a much
larger exponent on the order of 5 or greater. Such a value might be justified [35] by introducing a $\ll$thermal escape time$\gg$ for a VA pair out of their logarithmic
interaction potential into the generalized friction function determining the dynamic vortex correlator (see section 5.3). On the other hand it has been pointed out
[51,94] that finite size effects play in important role in determining $a(T)$.         \\

Shenoy [16] has developed a dynamic renormalization group approach for $JJAs$. The evaluation of the dynamic current-current correlation function
$F_c(\omega)$, defined in (4.16), is based on the
Fokker-Planck equation (4.1) for the probability distribution of the phases. Transforming the phase action (Hamiltonian (1.2) plus
friction term) to vortex variables allows borrowing results from static $BKT$-scaling and imposing it onto dynamics, by introducing an
extra frequency scaling. The dynamic conductivity $\sigma(\omega)$ of the array is then expressed as an integral over length scales $l$ of a scale
dependent conductance $\sigma(\omega,l)$. This approach was somewhat refined later on [107]. While the aproximations for $\sigma(\omega,l)$ in the first
calculation [16] essentially yielded a Drude-like frequency dependence, a more refined treatment of allowed to reproduce the $MP$ form
(5.6) in some intermediate frequency range, and thus to understand this anomalous behaviour from a more microscopic point of
view.

\section{ Disorder effects}

\subsection{Linear dynamics of disordered arrays}

In the considerations of section 5 the energy barrier that a moving vortex has to cross when moving from one plaquette of the array to a neighboring one (see
section 3.1) has been neglected. This approximation must be abandoned when the array deviates from perfect translational invariance. Different types of disorder can
be considered.  \\

     a. For ${\bf{bond~disorder}}$ the parameters characterizing the individual junctions, namely the Josphson couplings $J_{ll'}$, the junction resistances $R_{ll'}$ and the mutual
       capacities $C_{ll'}$, are random. This type of disorder is realized when the positions of the individual superconducting islands deviate in a random way from
       their ideal lattice positions or if they have random size, since $J_{ll'}$, $R_{ll'}$ and $C_{ll'}$ depend on the distance between the adjacent superconductors and on
       their geometry.       \\

      b. For ${\bf{site~ disorder}}$ the properties of the superconducting islands, ie their resistance and capacitance to ground, are random. More simply, in a $\ll$ dilute $\gg$
       array certain islands are totally missing.\\

Dynamic measurements have in particular been done on dilute, or $\ll$ percolative $\gg$, arrays in which only a fraction $p$ of superconducting islands is present [95]. The
experiments were done for $p$-values somewhat larger than the percolation threshold $p_c = 0.5$ where the disorder has a marked influence on the properties of the
array. Indeed for $p < p_c$ the average array has no more connected path joining opposite boundaries and superconductivity is completely suppressed. At temperatures
that are low compared to the $T_{BKT}$ (which in a dilute array is lower than for the regular counterpart, due to the absence of certain bonds, see, for example, section
17 of reference [1]) the equations of motion can be linearized, taking into account only small amplitude $\ll$spin wave$\gg$ excitations. The problem at hand is then
the same as determining the vibrational modes of a disordered harmonic solid, which is a well studied field [72]. In reference [73] the impedance $Z(\omega)$ of such a
dilute $JJA$ has been determined by evaluating the dynamic voltage correlation function (4.20) by different methods. The coherent potential approximation $(CPA)$ has
proven to be a useful approach for treating dynamic disorder problems. On the other hand Mori's method, applied in section 5.3 to vortex dynamics, is also useful.
The inverse of the impedance is expressed by an effective coupling function $K(\omega,p)$ depending on frequency and on the dilution fraction $p$ :
\begin{equation}
      Z(\omega)=(\eta+\frac{K(\omega,p)}{i\omega})^{-1}
\end{equation}
corresponding to the form (4.41) or to the effective inductance $L$ and resistance $R$ of (4.18).  \\

As a main result these calculations reveal the existence of three different frequency regimes:   \\

a) For very low frequency inductance and resistance are frequency independent. Such low frequencies effectively show the long wavelength response of the dilute
array. The details of the disordered structure are averaged out and the $\ll$macroscopic$\gg$ behaviour which is seen corresponds to a homogeneous superconductor.
However, $1/L$ and $R$ no more correspond to single junction parameters. They are reduced and go to zero when the percolation threshold $p_c$ is reached:
\begin{equation}
\begin{array}{rcl}
 \frac{1}{L(0,p)}\propto (p-p_c)^t\cr         \\
 R(0,p)\propto(p-p_c)^t
\end{array}
\end{equation}
where $t$ is the critical conductivity exponent introduced in percolation theory [74]. For the 2D lattice at hand $t = 1.3$, whereas the more simple $\ll$ mean field $\gg$
treatments like $CPA$ yield $t = 1$[73].                                  \\

b)  In an intermediate frequency range the dynamic conductance varies like a power of frequency :
\begin{equation}
   G(\omega)\sim \omega^{-u}
\end{equation}
reflecting dynamic scaling that results from the self-similar structure of the array near the percolation threshold. The precise value
of $u$ depends on the type of disorder and on the method (such as $CPA$ or Mori) used for the calculation. $CPA$ applied to bond disorder and
bond dissipation yields $u = 0.5$, which corresponds to the exact result found in percolation theory [75] by taking into account the
self-duality of the random network complex impedances by which the dilute $JJA$ can be modelled. On the other hand site dissipation yields
frequency exponents for $1/L$ and $R$ which differ from each other and are slightly smaller, respectively larger, than 0.5, their sum being,
however, equal to 1. The cross-over frequency where regime a goes over into regime b is itself a function of the dilution parameter $p$ :
\begin{equation}
  \omega_c(p)\propto(p-p_c)^{2t}
\end{equation}

This regime corrsponds to excitations which have a wave length smaller than the percolation correlation length and are therefore a direct witness of the disorder in
the array. In the field of disordered harmonic solids such modes are called $\ll$ fractons $\gg$.                      \\

c) At even higher frequencies $1/L$ and $R$ are again frequency independent and take the original values of a single junction multiplied by $p$. The dynamic response is
now governed by excitations of single bonds and the relevant parameters correspond to averages over the array.\\

Experimental data obtained on dilute arrays will be presented in section 8.

\subsection{Vortex dynamics of disordered arrays}

Deviations from a regular lattice structure will have an influence on the vortices of an array through the Peierls force or pinning potential in equation (3.5),
which has to be reintroduced in the equation of motion (3.6) or (3.9) for the vortex centers. The equilibrium arrangement of the (thermal or field induced) vortices
and antivortices will correspond to a minimum of the free energy in the given random pinning potential landscape. It is generally accepted that weak disorder
results in some sort of glassy vortex state [76,77]. At sufficiently low temperatures such a vortex configuration would then be trapped by the underlying
potential, and the system should have a vanishing linear resistance, ie it would be truly superconducting. This should be valid for dilute JJAs, as they have been
described in the preceding subsection, as well as for ultrathin superconducting films [78] that are intrinsically disordered. There is, however, an important
difference between 2D and 3D systems : in two dimensions the V-system has thermally excited defects, in particular dislocation pairs, which allow for local flux
flow (and thus ohmic resistance), even though any motion of the V-A-structure as a whole is quenched by the pinning potential. Such a motion can, however, be very
slow, so that on the time scale of a given experiment the V-system still appears to be frozen. Recent measurements showing such a « $\ll$dynamic freezing$\gg$ » will be
presented in section 8 [78].      \\

In the percolative arrays described in the preceding subsection the sign of disorder has also been detected in the vortex dynamics. Above the BKT transition the
vortex contribution to the dynamic impedance $Z(\omega)$ shows power law behaviour with a temperature dependent exponent. These data [98] will be interpreted at the end of
subsection 8.2.

\section{Theoretical description of experiments probing equilibrium dynamics}

Three major experimental techniques have been developed in order to get insight into the dynamic behaviour of $JJA's$  at or near equilibrium: \\

a) I-V characteristics (V-response to static driving fields)    \\

b) Flux noise (spontaneous V fluctuations)                            \\

c) Two-coil induction (V-response to weak dynamic driving fields)\\

Before reporting on experimental findings we briefly explain the link between the measured quantities and the theoretical techniques exposed in section 5.

\subsection{Current-voltage characteristics}

Below the $BKT$ transition there are no free vortices and the dc resistance is zero. However, an applied current will act on the vortices according to equation
(3.6) which leads to a finite probability for a pair to unbind and thus to dissipation. Experimentally this phenomenon can be observed through the non-linear
current-voltage characteristics 
\begin{equation}
           V\propto I^{a(T)}
\end{equation}

The form of the temperature dependent exponent $a(T)$ for $T < T_{BKT}$ can be found setting the escape rate over the energy barrier for a pair equal to the probability
for recombination. Details can be found in the original work of Ambegaokar et al [34] or in the review by Newrock et al [1]. The result was given in (5.23). Thus,
for $T < T_{BKT}$, the I-V relation is non-ohmic, while above the unbinding temperature the presence of free vortices leads to ohmic dissipation.  \\

As mentioned in section 5.4 dynamic scaling leads to result (5.24) which is different except at $T = T_{BKT}$. Bormann [70] has refined and extended the picture of this
balance between pair breaking and recombination. At not too low densities of vortices charge transport is supposed to occur predominantly through direct particle
transfer from one bound pair to a neighboring one. Taking into account such $\ll$ partner transfer $\gg$ Bormann indeed obtains the I-V exponent of [49] thus providing a
more physical insight than simply dynamic scaling. However, in contrast to reference [49] Bormann's result for the exponent $a(T)$ goes over into the $\ll$ classial $\gg$
form proposed in [34] for very low currents. Thus the apparent contradiction between the two forms of [34] and [49] for $a(T)$ is resolved.  \\

Theoretical values for $a(T)$ resulting from numerical simulations will be presented in section 8 and experimental findings for the I-V characteristics of $JJAs$ will
be shown in section 9.  For an adequate comparison between theory and experiment one has to bear in mind that finite size effects can influence the $BKT$ scenario in
particular through the presence of some free vortices which are not the result of $BKT$ unbinding, but rather of the finiteness of the system. Such finite size
effects have been discussed by different authors, see [51], [71],[79]. An observable effect of this phenomenon will be discussed in section 8.1.

\subsection{Two-coil induction}

These measurements use a drive coil of $N_D$ turns of radius $R_D$ and a pair of astatically wound receive coils of $N_R$ turns of a smaller radius $R_R$ mounted at some small
distance $d$ from the plane of the array (see $figure~ 2$). In order to study the dynamic response of the array an ac magnetic field is created by the drive coil, which excites the
vortex system. The resulting motion of the vortices creates a voltage signal $\delta V$ in the receive coil. This technique was pioneered by Hebard and Fiory [80] for
superconducting films and has proven to be very successful for studying the dynamic aspects of the $BKT$ transition [1,81-85].    \\

\vspace{10cm}

    $Figure~ 2 ~:$Two-coil set-up used in the measurement of the dynamic conductance of the array, described in section 7.2.              \\

The measured signal $\delta V$ can be related to the dynamic conductance of the array by a straightforward calculation. Since the latter can be found in various references
[81,83], we only sketch the main steps without giving an explicit derivation :            \\
- An ac current $I_D(\omega)$ of frequency $\omega$ sent through the drive coil produces a time dependent vector potential ${\bf{A}}(\omega)$  \\
- The time derivative of ${\bf{A}}$, in the plane of the array, is an electric field ${\bf{E}}(\omega)$ acting on the latter   \\
- An electric current density ${\bf{j}}(\omega) = G(\omega){\bf{E}}(\omega)$ is induced in the array  that acts on the vortices     \\
- The resulting  time dependent flux threads through the receive coil and can be detected by the induced voltage $\delta V(\omega)$.   \\

Thus the signal $\delta V(\omega)$ is linearly related to the drive current $I_D$:
\begin{equation}
   \delta V(\omega)=i\omega I_D(\omega)\int_0^{\infty}dx\frac{M(x)}{1+\frac{2x}{\mu_oh}\frac{1}{i\omega G(\omega)}}
\end{equation}

Here $h$ is the sum of the closest distances of the drive and the receive coils, respectively, from the array plane and the function $M(x)$ takes into account the
geometry of the experimental set-up 
\begin{equation}
    M(x)=\pi\mu_o\frac{R_DR_R}{h^2}J_1(\frac{R_D}{h}x) J_1(\frac{R_R}{h}x)e^{-x}\frac{1-e^{-N_D\delta h_Dx}}{1-e^{-\delta h_Dx}}
\end{equation}
$\delta h_D, ~\delta h_R$ being, respectively, the distance between the turns of the drive and the receive coils, in units of $h$ and $J_1$ is the first
order Bessel function. In order to bring in vortex dynamics into the expression (7.2) one then has to use a link between the electric conductance $G$ of the array and the vortex dielectric function $\epsilon(\omega)$, such as (4.45). More details concerning the
derivation of (7.2) and (7.3) can be found, for example, in [83].\\

Even though finite frequencies are involved this technique allows to gain interesting experimental insight into the $BKT$-transition. Moreover it has been used to
detect the frequency dependence of $\epsilon(\omega)$. Results will be presented in section 9.

\subsection{Flux noise }

Currents flowing in the array produce a magnetic field that extends into the space above and below the plane of the array. This field can be detected by measuring
the corresponding magnetic flux through a given surface. In the usual experiment a circular wire of radius $R$ delimiting such a closed surface is fixed at a distance
$d$ from the $JJA$. If the field, and thus the flux, varies with time a voltage is induced in the wire according to Faraday's law. In the absence of an external
excitation (which was used in the two-coil experiment) the time variation of the flux is due to thermal fluctuations in the array. Such a $\ll$flux noise$\gg$ experiment
measures $S_{\phi}(\omega)$, the frequency Fourier transform of $S_{\phi}(t)$, which is the dynamic correlation function $\langle\phi(t)\phi(0)\rangle$ of the time
dependent flux $\phi(t)$ through the area surrounded by the wire.\\

The usual theoretical analysis of flux noise attributes the fluctuating currents in the array to moving vortices. The flux $\phi$ is expressed as an integral over the dynamic
charge density correlator of the vortex system in the plane by means of the following steps (we give some more details here since the relevant references do not
give a complete derivation):   \\

- $\phi$ is the line integral of the vector potential ${\bf{A}}$ along the wire
\begin{equation}
  \phi=\oint d{\bf{s}}\cdot {\bf{A}}
\end{equation}

- The vector potential ${\bf{A}}$ which is produced by the electric current density ${\bf{j}}$ flowing in the array is found by solving Maxwell's equation
\begin{equation}
{\bf{A}}({\bf{r}})=\mu_o\int d^2k_{\parallel}e^{i{\bf{k}}_{\parallel}\cdot{\bf{r}}_{\parallel}} \frac{e^{-k_{\parallel}r_{\parallel}}}{2k_{\parallel}}{\bf{K}}_{in}
({\bf{k}}_{\parallel})
\end{equation}

Here the position vector ${\bf{r}} = ( {\bf{r}}_{\parallel}, {\bf{r}}_{\perp})$ is decomposed into its components in the plane of the array and
perpendicular to the latter and ${\bf{K}}_{in}$ is the 2-dimensional Fourier
transform of the areal current density in the array. 

- For a single vortex the current ${\bf{K}}$, circulating around its center, is given by
\begin{equation}
{\bf{K}}({\bf{r}}_{\parallel})=\frac{2e}{\hbar}J(\nabla\theta({\bf{r}}_{\parallel})-\frac{2e}{\hbar}{\bf{A}}({\bf{r}}_{\parallel}))
\end{equation}

Equations (7.5) and (7.6) have to be taken into account simultaneously which leads to magnetic screening  in the final expression for the vector potential:
 \begin{equation}
             {\bf{A}}( {\bf{r}})=\mu_o   \int d^2k_{\parallel} e^{i {\bf{k}}_{\parallel} \cdot  {\bf{r}}_{\parallel}   }\frac{e^{-k_{\parallel}r_{\parallel}}}{2k_{\parallel}}
             \frac{{\bf{k}}_{in}({\bf{k}}_{\parallel})}{1+\frac{1}{\lambda_{\parallel}k_{\parallel}}}
  \end{equation}
The effective (in-plane) screening length $\lambda_{\parallel}$ is given by
\begin{displaymath}
  \lambda_{\parallel}=\frac{\hbar^2}{2e^2J\mu_o}
\end{displaymath}
- Assuming that the relevant contributions to the phase gradient are due to moving vortices the current density ${\bf{K}}$ is expressed by the Fourier transform of the
density of the latter : 
\begin{equation}
 {\bf{K}}_{in}({\bf{k}}_{\parallel})=-2\pi i(\frac{2e}{\hbar})J\frac{{\bf{e}}_{\perp}\times{\bf{k}}_{\parallel}}{k_{\parallel}^2} \rho_v({\bf{k}}_{\parallel})
\end{equation}

- Introducing expressions (7.5) to (7.8) into (7.4) leads to an explicit expression of the flux $\phi$ through the plane of the wire in terms of the vortex density in the plane of
the array : 
\begin{equation}
     \phi=\pi R\mu_o(-2\pi i)\frac{2e}{\hbar}J\int_0^{\infty}dk_{\parallel}k_{\parallel}J_1(k_{\parallel}R)e^{-k_{\parallel}d}\frac{\lambda_{\parallel}k_{\parallel}}{1+
     \lambda_{\parallel}k_{\parallel}}\frac{\rho_v(k_{\parallel})}{k_{\parallel}^2}
\end{equation}
where $J_1$ is the first order Bessel function bringing in the geometry of the pick up wire.\\

- The flux noise correlation is given by
\begin{equation}
   S_{\phi}(\omega)=\int dt~e^{i\omega t}\langle\phi(t)\phi(0)\rangle
\end{equation}

- The statistical average concerns the vortex density showing up in statement (7.9). It leads to the dynamic density correlator of the V-A system introduced in
equation (4.24). The latter is diagonal in the wave vector $k$ (as it is defined in (4.24)). Thus the $S_{\phi}(\omega)$ is expressed by the Fourier transform of the
dynamic vortex correlation function which, in turn, is related to the Laplace transform (4.24) by
\begin{equation}
  \int dt~e^{i\omega t}\langle\rho_v(k,t)\rho_v^*(k,0)\rangle=2Re\phi_{\rho\rho}(k,-i\omega)
\end{equation}

- The final result for the flux noise spectrum reads 

\begin{equation}
    S_{\phi}(\omega)=(2\pi ^2R\mu_o)^2(\frac{2e}{\hbar})^2(J\lambda_{\parallel})^2\int_0^{\infty}dk\frac{J_1(kR)^2e^{-2kd}}{k(1+\lambda_{\parallel}k)^2}2Re\phi_{\rho\rho}(k,-i\omega)
\end{equation}

In case several concentric wires are set up on top of each other (7.12) has to be replaced by a sum of terms of the same structure, but each with the corresponding
distance $d_n$ from the array, showing up in the exponent. Results obtained through expression (7.10) by using different forms for the dynamic vortex correlation
function $\phi_{\rho\rho}(p,z)$ will be shown in section 8. They will be compared with experimental data in section 9.    \\

In order to build a bridge between flux noise and electrical conductance one can use the links between charge density correlator, dynamic susceptibility and
dielectric function shown in (4.26,27) in order to rewrite (7.12) as
 \begin{equation}
    S_{\phi}(\omega)=(2\pi ^2R\mu_o)^2(\frac{2e}{\hbar})^2(J\lambda_{\parallel})^2\int_0^{\infty}dk\frac{J_1(kR)^2e^{-2kd}}{k(1+\lambda_{\parallel}k)^2}\frac{2}{\omega}\frac{k_BTk^2}{2\pi q_o^2n} \frac{1}{\epsilon_v(\omega)}
\end{equation}

When the pick-up coil is very close to the plane of the array ($d$ small) only small $k$-values contribute to (7.11) which can be approximated by
\begin{equation}
    S_{\phi}(\omega)=C\frac{1}{\omega}Im(\frac{1}{\epsilon_v(\omega)})  =C'ReG_s(\omega)
\end{equation}
relating thus - through (4.45) - flux noise approximately to the long wave length conductivity, or - more precisely - to its superconducting part [48].  \\

The above flux noise calculations have been based on taking into account only the supercurrent in the array (see equation (7.6)) and by relating the phase gradient
in (7.6) directly to vortex configurations. The link between flux noise and conductance can be made more general [86] by expressing first the flux correlator
(7.10) by the correlator of the total array current and by relating the latter to the array conductance via the considerations exposed in section 4 on linear
response.

\section{ Experimental data}

We present experimental data for the three types of $\ll$standard$\gg$ experiments for which we have related the measured observables to theoretically accessible
quantities in the preceding section.

\subsection{Current-voltage characteristics}

Various measurements of the relationship between an applied current and the resulting voltage drop, yielding the exponent $a(T)$ in (7.1), are cited in the review by
Newrock et al [1]. There (in section 11) the interplay between different relevant length scales implied in such an experiment, where currents of different
intensity are used, is also explained in detail. The jump from the non-ohmic behaviour given by a value of $a(T)$ that differs from 1 to the high temperature linear
I-V relation is clearly bourne out by experiment. \\

As explained in sections 4 and 6 a new $T$-dependence [49] of the coefficient $a$, differing from the $\ll$ classical $\gg$ form has recently been proposed. Weber et al
[64] have re-analysed I-V data (see $Figure~3$) obtained on Hg-Xe alloy films [88], $Bi_2Sr_2CaCu_2O_x $[89] and $Bi_{1.6}Pb_{0.4}Sr_2Ca_2Cu_3O_x$ [90] single-crystal films, finding indeed
good agreement with the new form of $a(T)$ of Ref. [49] and [70]. Herbert et al [79] have come to the same conclusion for proximity-coupled and high-$T_c$ weak-link
arrays. They also find evidence for the finite size effects mentioned at the end of section 7.1. Indeed, the I-V curves for very weak currents bend over into a
linear form even below the transition temperature, since the free vortices present in a finite array, lead to ohmic behaviour.      \\

\vspace{8.5 cm}

  $Figure~ 3 :$ Exponent $a(T)$ of the I-V characteristics (see equation (7.1)), according to Ref. 87, plotted as a function of
$X=T/T_c$. The different curves have the following meaning :        \\
- Plusses : Reference 88\\
- Other symbols : Monte Carlo results for an array of length of 16 (stars), 24 (open circles), 32 (filled circles) and 48 (triangles) unit cells.  \\
The dashed line corresponds to the universal value $a=2$, at the critical temperature. \\
The inset shows experimental data for $Hg-Xe$ alloy films, Ref. 88 (plusses), $Bi_2Sr_2CaCu_2O_x$ single-crystal films,
Ref. 89 (filled squares) and $Bi_{1.6}Pb_{0.4}Sr_2Ca_2Cu_3O_x$ single-crystal films, ref. 90 (open squares). \\

\subsection{Two-coil induction}

The two-coil mutual induction technique, from which the dynamic conductance $G(\omega)$ of the array can be extracted (see section 7), has extensively been used in order
to study the $BKT$ transition. One of the basic features is the temperature dependence of the phase stiffness or the helicity modulus $\land$, reaching a universal value at
$T_{BKT}$ before jumping to zero. As explained in section 4 the low frequency limit of $G(\omega)$ is directly related to $\land$. Ref. [81] presents data obtained on a
proximity-effect $JJA$ consisting of Pb-islands on a Cu substrate. The voltage induced in the pick-up coil is analysed according to section 7.2 and the corresponding
dynamic dielectric function $\epsilon(\omega)$ is extracted for different temperatures in the vicinity of $T_{BKT}$. The fall-off of $Re(\epsilon^{-1})$ near $T_{BKT}$ is characteristic of the
$\ll$insulator-conductor transition$\gg$ of the vortex system and $Im(\epsilon^{-1})$ shows that this transition is accompanied by dissipation. Going to
higher frequency shifts the transition region to higher $T$, since larger $\omega$ tests shorter length scales for which the unbinding of V-A pairs happens at higher temperature
$T_{\omega}$, which can be
interpreted as an $\ll$ effective frequency dependent critical temperature $\gg$. In order to extract the helicity modulus $\land$ which is a thermodynamic response coefficient
one has to extrapolate to zero frequency. This can be done by identifying $Re(\epsilon(\omega,T))$ for a given frequency $\omega$ with the static dielectric function
$\epsilon(l_{\omega},T)$ for a length scale $l_{\omega}$ corresponding to $\omega$. The $T$-dependence of $\land$ obtained in this way is in good agreement with a renormalization group calculation integrating the scaling
equations up to a length of the order of $\xi(T_{\omega})$, the $BKT$ correlation length at $T = T_{\omega}$ where pairs of length $\xi(T_{\omega})$ begin to dissociate. The extrapolation to infinite
length scale agrees very well with the universal jump of $ \land $ at $T_{BKT}$ prredicted by theory. More details concerning this scaling procedure used for
 getting at $\land $ can be found in [81] and [91].    \\

It is important to mention that whenever experiment and theory is compared in this way one has to take into account the $T$-dependence of the Josephson coupling
given, according to (1.3), by the one of the critical current
\begin{equation}
   I_c(T)=I_o(1-\frac{T}{T_{BCS}})^2exp(-\frac{L}{\xi_n(T)})
\end{equation}
see [92], where $T_{BCS}$ is the $BCS$ transition temperature of the individual islands of the array, separated by the distance $L$ and $\xi_n$ is the effective coherence
length of the normal metal forming the junction between two neighboring islands. One can then compare experimental data with $XY$-model calculations by using a
dimensionless temperature parameter 
\begin{equation}
     \hat{T}=\frac{k_BT}{J(T)}=\frac{2ek_BT}{\hbar I_c(T)}
\end{equation}

Moreover, spin wave excitations produce an additional renormalization of the Josephson coupling. The latter can be taken into account by using an $\ll$ effective
Coulomb gas temperature $\gg$ [93]
\begin{equation}
  \tau=\frac{\hat{T}}{1-\hat{T}/4}
\end{equation}

The two-coil technique has also allowed to shed light on the dynamics of 2D Coulomb systems. In section 5 two phenomenological approaches to $\epsilon(\omega)$ have been
presented : Drude's model and Minnhagen's phenomenology $(MP)$. One might expect that above $T_{BKT}$, where free vortices are supposed to dominate the response to a
driving force, Drude's model would be appropriate. It thus came as a surprise when measurements on weakly frustrated triangular $JJAs$ [82] revealed a frequency
dependence of $\epsilon(\omega)$, see $figure~4$, which is well described by $MP$ and support the idea of a non-analytic dynamic mobility, as explained in section 5.2. Triangular arrays are better
suited for studying the intrinsic dynamics of interacting vortices since the barriers due to lattice pinning (section 3.1) are very low [23]. One has to bear in
mind that the system of field induced vortices in a frustrated array should actually be described as a $\ll$ one-component plasma in a neutralizing background $\gg$
($OCP$) rather than the $\ll$neutral two-component plasma$\gg$ ($TCP$) of the thermally excited V-A system in an unfrustrated array. Thus, theoretical approaches developed for
the $TCP$, as in section 5, are not immediately applicable. However, numerical simulations of frustrated arrays (section 9) have shown that the two types of plasmas
behave the same way.         \\

\vspace{8.5cm}

  $Figure ~4:$ Real and imaginary parts of the inverse of the vortex dielectric function measured in reference 82 on a $JJA$ in a
small magnetic field. The two functions are plotted as a function of the reciprocal frequency. The full lines are
fits to the data using Minnhagen's expressions (5.6), whereas the Drude form is given by the dashed lines.        \\

It is clear that such a vortex mobility of the form (5.6) cannot hold down to arbitrarily low frequency above the vortex unbinding transition. Indeed, the
electrical conductance $G(\omega)$ is related to the vortex mobility according to (4.46). The vanishing vortex mobility (5.6) for $\omega\to 0$ of MP would therefore imply that
superconductivity is maintained even above $T_{BKT}$.                                     \\

Another interesting application of the two-coil technique has shed light on the dynamics of percolative arrays [95] discussed theoretically in section 6, see $figure~ 5$. The data
confirm that the transition termperature goes to a lower value when the concentration of missing sites increases, ie when the percolation threshold is approached.
Moreover, both the inverse inductance and the resistance show very clearly two of the three frequency regimes discussed in section 6 : for low frequencies both
quantities have constant values given by (6.2), whereas above some cross-over frequency the fractal behaviour is observed. The critical exponent $t \approx 1.4$ is
extracted from the data which is in good agreement with the prediction $t \approx 1.3$ for percolation in two dimensions [96]. The frequency dependence in the fractal
regime is compatible with $u\approx 0.55$, see equation (6.3) which is somewhat larger than what is expected from duality arguments for bond percolation [75] and for a mean field
approximation. On the other hand, different types of approximations used in the theoretical description of disordered arrays in Ref. [73] also lead to u values
that differ from 0.5.    \\

       \vspace{8cm}

       $Figure~ 5 :$ Log-log plot of the frequency dependence of the inverse kinetic inductance $L^{-1}$ and resistance $R$ (according to
equation (4.18)) of a dilute array close to the percolative limit $(p = 0.51)$ from reference 95. The dashed lines
indicate $\omega^{1/2}$ power laws.\\

 Dynamic response measurements on the dilute arrays near their perconlation threshold have also been extended to temperatures near and above the expected $BKT$
temperature [98], where vortex motion is expected to dominate over the linear spin wave dynamics discussed before. The contribution $Z_v(\omega)$ of the thermally excited
V-A-system to the impedance $Z(\omega)$ is extracted from the measured values by representing the impedance of the array, in the spirit of the $RSJ$ model, by two parallel
channels - an inductive one for $\ll$ superfluid $\gg$ conduction and a dissipative Ohmic path - and by assuming that the vortex response renormalizes the superfluid part.
A similar decomposition as (4.18) fo the dynamic conductance is used for $Z_v$
 \begin{equation}
          Z_v(\omega)=R_v(\omega)+i\omega L_v(\omega)
\end{equation}
(whereby one has to bear in mind that resistances and inductances introduced, either by (4.18) or by (8.4) are not identical ). Above $T_{BKT}$ both, $R_v$ and $L_v$, show a
power law behaviour with frequency
 \begin{equation}
       R_v(\omega)\propto \omega^{u(T)}~,~L_v(\omega)\propto \omega^{u(T)-1}
\end{equation}
with an temperature exponent. The latter is close to 1 at $T_{BKT}$ and decreases slowly when $T$ increases. Such a $T-$dependent exponent can be obtained by assuming that
the relevant contribution to the dynamic response comes from pairs and by calculating their polarisability as indicated in subsection 5.1. However, a simple
calculation using statement (5.3) yields an exponent $u(T)$ that decreases much faster with rising $T$ than the ones observed [98]. Thus, one should probably revise
the ideas of subsection 5.1 for disordered arrays. The response at temperature much higher than $T_{BKT}$ again points to the motion of free vortices, the latter being
activated across the pinning barriers.        \\

Glassy behaviour of the vortex system, mentioned in section 6.2, has been seen in measurements of the dynamic impedance $Z(\omega)$ in ultrathin $YBa_2Cu_3O_7 (YBCO)$ films in
a perpendicular magnetic field [78]. At high enough temperatures the dissipative component of $Z$ has an activated $T$-dependence resulting from activated motion of single
(field induced) vortices out of local pinning wells. With decreasing temperature $Z$ crosses over to a more or less $T$ independent value. The cross-over temperature
$T^*$, that depends weakly on $\omega$, turns out to be much higher than a reasonable estimate for the melting temperature of a 2D V-lattice, due to unbinding of dislocation
pairs. This points to the effect of pinning on the V-system. The frequency dependent response of such a pinned elastic manifold has been studied theoretically by
various authors [99]. The corresponding predictions agree well with the measurements on $YBCO$ films. It would, of course, be interesting to see in more refined
experiments whether $T^*(\omega)$ decreases without limit when the frequency $\omega$ becomes lower and lower. This would be a true sign of the fact that the 2D $\ll$ glassyness $\gg$ is
really a dynamic phenomenon.\\

\subsection{Flux noise}

The flux noise spectrum $S_{\phi}(\omega)$ has been measured in overdamped $JJAs$ by Shaw et al [68] and more recently by Candia et al [100], as well as in high-$T_c (YBCO)$ films
by Festin et al [101]. Shaw's data cover the range from $T = 1.825 K$ to $T = 2.379 K$ for a $Nb$ on $Cu$ array with $T_{BKT} \approx 1.63 K$, see $figure ~6$. At each $T,~ S_{\phi}(\omega)$
is white for $\omega < \omega_c$ and varies like $1/\omega$ for $\omega\ge\omega_c$ . The cross-over frequency $\omega_c$ is proportional to the inverse
square of the Kosterlitz-Thouless correlation length with the $T$ dependence
as given by (1.12). This result is consistent with dynamic scaling (see section 5.5), for a dynamic exponent $z = 2$ and with $b\approx 2$ in expression (1.12). The white noise level increases
for decreasing $T$ as $\xi_{BKT}(T)^2$. The authors conclude that, even at the lowest temperature shown (ie., for $T$ as close to $T_{BKT}$ as possible), $\xi_{BKT}$ is on the order of the
size of the pick-up coil and, as a consequence, the prefactor $\xi_o$ in (1.12) is actually smaller than the lattice constant of the array. The fact that scaling is obeyed down to
such a microscopic length scale is somewhat surprising.\\

The data of ref. [101] confirm previous observations [102] showing that high-$T_c$ films also show typical 2D vortex fluctuations. Data are shown for 500 A thick
$YBCO$ films from $T = 84.84 K$ to $T = 84.89 K$, just slightly above $T_c$. At low frequency $S_{\phi}(\omega)$ is again white, near some frequency $\omega_o$ the data have a common tangent
corresponding to  
\begin{equation}
        S_{\phi}(\omega)\sim\omega^{-1.05}
\end{equation}
before crossing over to $1/\omega^{3/2}$. However, contraray to what has been seen by Shaw et al [68] the noise spectrum does not have a true $1/\omega$ dependence over any
appreciable frequency range. Thus, the two materials ($JJA$, respectively high $T_c$ film) have a different characteristic behaviour : while the $JJA$ shows an impressive
$1/\omega$ noise, the film data only have a $1/\omega$ tangent whereas the high frequency behavior is $1/\omega^{3/2}$ .
The same is found in [102] for $Bi_2Sr_2CaCu_2O_{8+x}$ films.\\
\pagebreak

\begin{figure} [h]
\let\picnaturalsize=N
\def\picsize{8 cm}
\def\picfilename{rawdata.eps}
\ifx\nopictures Y\else{\ifx\epsfloaded Y\else\input epsf \fi
\let\epsfloaded=Y
\centerline{\ifx\picnaturalsize N\epsfxsize \picsize\fi \epsfbox{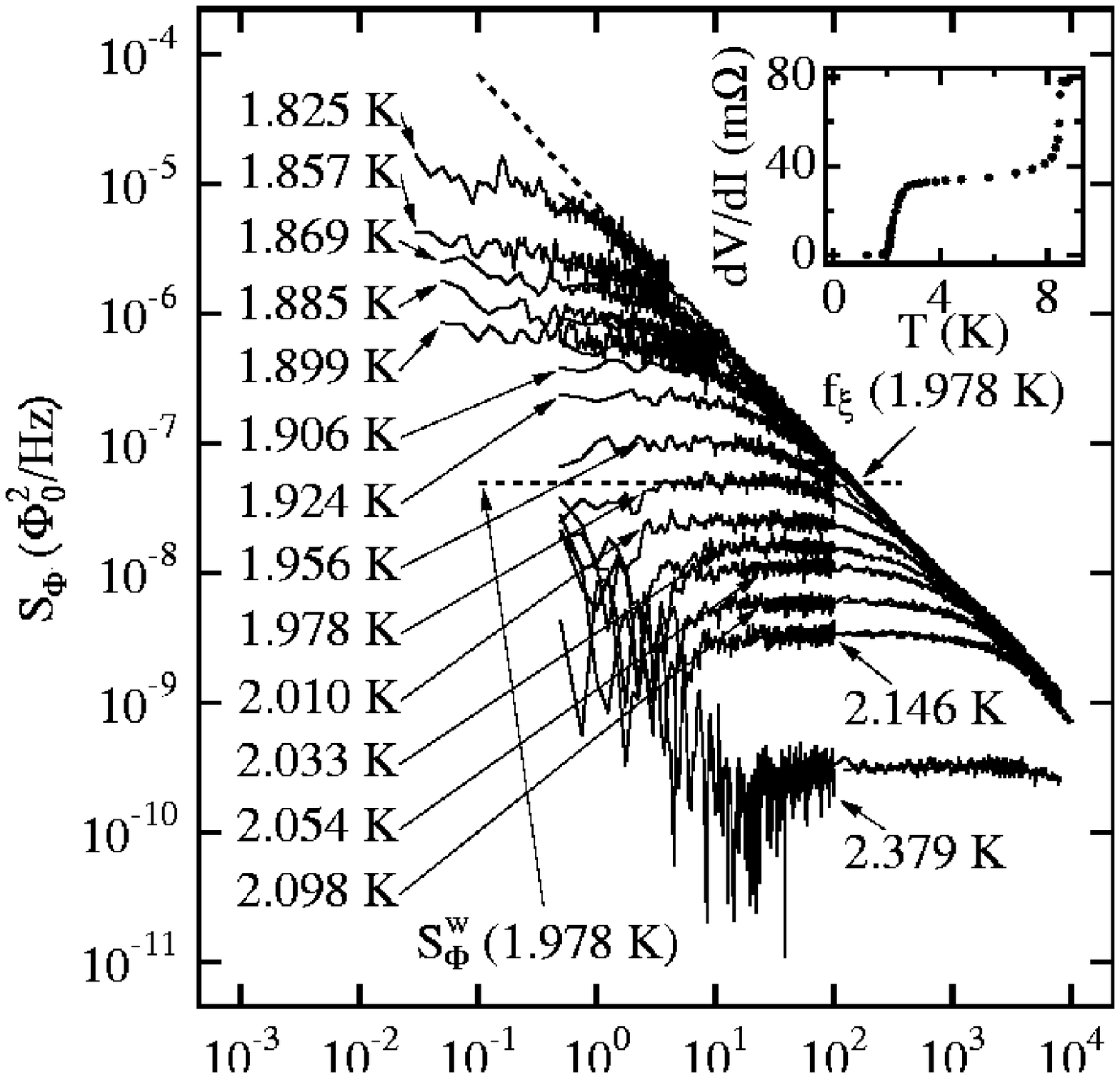}}}\fi
\label{AmplitudePhi4}
\end{figure}

$Figure~ 6:$ Flux noise spectrum versus frequency for different temperatures of a $JJA$ consisting of $Nb$ islands on $Cu$ (ref.
68).The dashed lines have slope -1 and 0, respectively, signalling white noise and $1/\omega$ noise in different
frequency regimes. The inset shows the derivative of the current voltage characteristics, $dV/dI$, as a function of
temperature. The $BKT$ temperature of the array is at $1.63 K$.\\

The most recent measurements on a proximity effect $Pb/Cu/Pb$ triangular $JJA$ [100] confirm the $1/\omega$-dependence of $S_{\phi}$ above $T_{BKT}$ (see $figure~ 7$), but the analysis of the low frequency
white noise level has been interpreted in terms of an exponentially activated (total) number $n(T)$ of vortex excitations. This would show that the vortex mobility
determining (7.13) through (4.28), (see also (9.5)),
which is supposed to go to zero when the critical temperature is approached, does not really influence the white noise level. This may be a
finite size effect (see section 9.3 where results of theoretical flux noise calculations will be presented). Candia et al [100] have also extended their
measurements to temperatures close to, but below $T_{BKT}$. In the frequency range covered by the measurements the noise keeps to behave like $1/ \omega$ but, for a given
frequency, the level decreases when $T$ goes down.       \\

The ubiquity of the $1/\omega$-dependence is somewhat surprising. As shown in section 9 it can be reproduced when pair dynamics is taken into account, but with an exponent
in the probability distribution for the relevant pair distances in statement (5.2) which is frozen at $T = T_{BKT}$. Thus it seems that the array $\ll$ stays critical $\gg$
even when the temperature differes from its critical value, as it has been remarked by Shaw et al [68]. In the context of disordered arrays [98], the
corresponding exponent has also been extracted from dynamic data and shown to be much less temperature dependent than statement (5.2) would imply. As a matter of
fact, the $1/\omega$ dependence could quite generally be a disorder effect. Indeed, even small variations of the distance $d$ between neighboring superconducting islands
implies a relatively strong $\ll$ disorder $\gg$ in the Josephson coupling constants, since the latter depend exponentially on $d$. In such a disordered array the free
vortices that can be present at any temperature will have to move on a $\ll$ random substrate $\gg$. The corresponding dynamic response would typically behave like $1/\omega$.

\pagebreak

     \begin{figure} [h]
\let\picnaturalsize=N
\def\picsize{8 cm}
\def\picfilename{fig2.eps}
\ifx\nopictures Y\else{\ifx\epsfloaded Y\else\input epsf \fi
\let\epsfloaded=Y
\centerline{\ifx\picnaturalsize N\epsfxsize \picsize\fi \epsfbox{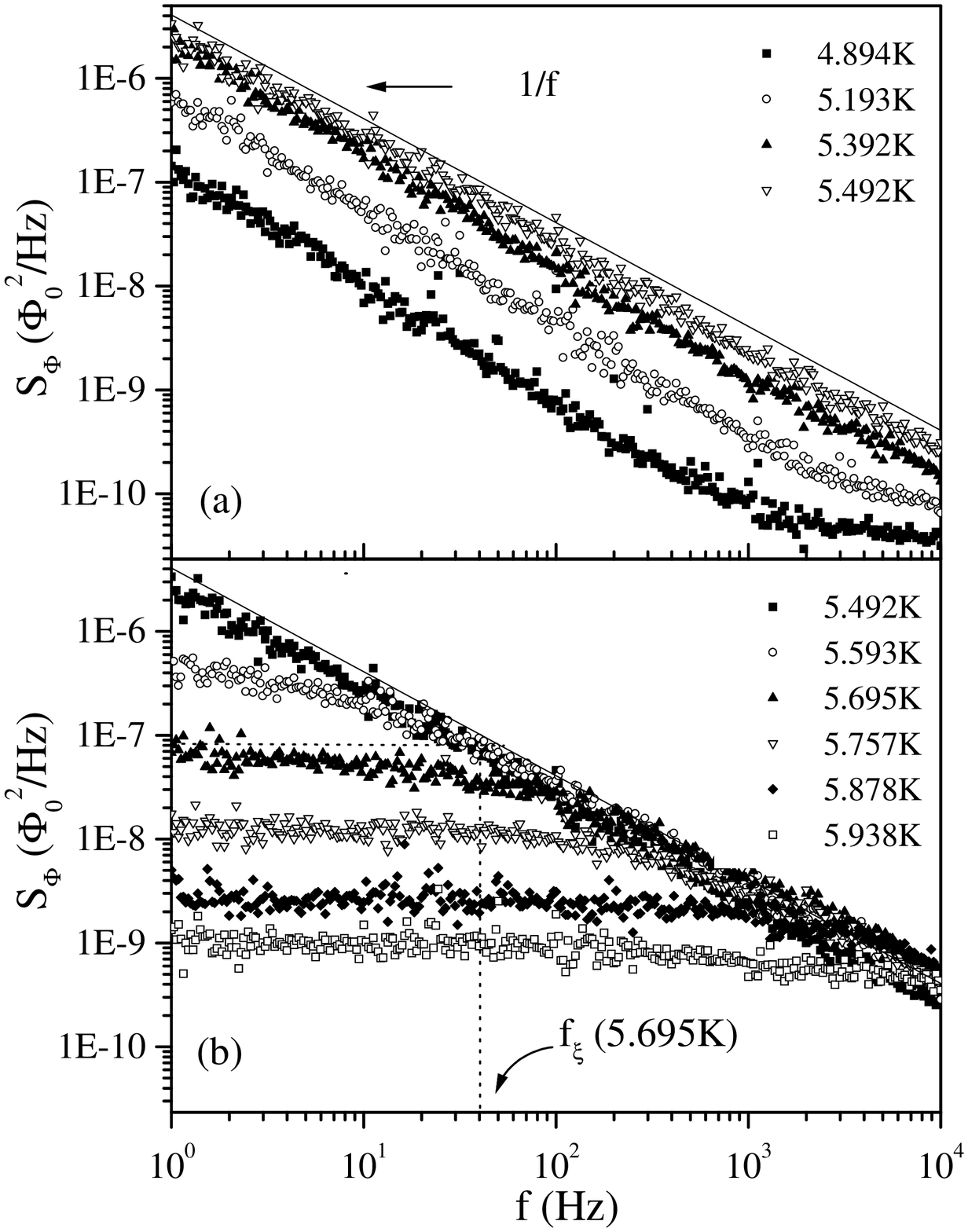}}}\fi
\label{AmplitudePhi4}
\end{figure}

$Figure~ 7 :$ Flux noise spectrum from reference 100, measured on a $Pb/Cu$ array as a function of frequency $f$ at temperatures
slightly below (a) and slightly above (b) the $BKT$ transition.\\

     \section{Theoretical results}

 \subsection{Dynamic critical exponents}

The dynamic exponent $z$, introduced and discussed in section 5.5, has been determined through various numerical simulations. Tiesinga et al [46] solved the
equations of motion for the phase field using alternatively $TDGL$ and $RSJ$ dynamics, as explained in section 5.4. The corresponding value of $z$ is obtained by studying
the scaling form (5.21) of their correlation functions, ie by identifying the link between the relevant relaxation frequency and the $BKT$ correlation length.
Interestingly, they find that $z$ is different for the two types of dynamics : $z(TDGL) \approx 2$ corresponds to what is expected, whereas $z(RSJ) \approx 0.9$
 points to a different type of critical dynamics(see the insets in figure 10 that will be discussed in connection with flux noise). This would be an example of a system with a given static critical behaivour ($BKT$ in this case) which goes over into
 different dynamic universality classes, depending on the precise equations of motion used to describe time evolution.         \\

The scaling analysis of the $TDGL$ and $RSJ$ simulations of Kim et al [53], who used periodic and fluctuating twist boundary conditions ($FTBC$, see section 5.4),
yields $z=2$ at the transition temperature for both types of dynamics for $FTBC$. This value is found by finite size scaling (extrapolation of calculated parameter
values to the thermodynamic limit) of the array resistance. However, they observe that finite size scaling with $PBC$ gave different results, or - more precisely -
that $PBC$ are actually inadequate for this procedure, since the relaxation time which reveals the exponent $z$ is proportional to the macroscopic resistance that is
actually zero for periodic boundary conditions. This shows once again that correct boundary conditions are crucial whenever transport coefficients are considered,
which describe the response of the system to an external influence that couples to the boundary. Below $T_{BKT}$ the value of $z$ is slightly larger than 2, which is in
agreement with the finding for a 2-d lattice Coulomb gas with Monte Carlo dynamics [64].

           \subsection{Vortex dielectric function}

The fact that the vortex dielectric function, measured by the two-coil technique described in the previous section, has a frequency dependence resembling more the
$\ll$Minnhagen phenomenology$\gg$ ($MP$) than the classical Drude behaviour, has initiated both analytical and numerical work aiming at understanding this
 $\ll$anomalous$\gg$ dynamic response of $JJAs$. Such an effect is likely to be related to vortex-vortex interaction, to pinning by the underlying lattice or to the fact that vortices are
not just point particles, but singular excitations of the phase field which change shape when they move and when they interact with each other.  \\

Coupling to spin waves [103,104], which are the $\ll$Bosons$\gg$ mediating the Coulomb interaction between vortices in the framework of the effective electrodynamics
described in section 4.2, has been shown to lead to a vortex dielectric function of the form given in section 5.2. The non-analytic low frequency dependence, which
is characteristic of $MP$, is an $\ll$ infrared divergence $\gg$ that can be understood by counting powers of wave number in the relevant integrals. It is the combined effect
of the spin wave propagator and the dimensionality of the array. Other authors [105,106] have also found non-analytic effects in vortex dynamics, for instance in
the vortex mass of arrays with capacitive effects, caused by coupling to spin waves.  \\

Capezzali et al [107] have treated the problem of interacting vortices, both by a refined dynamic scaling (a generalization of the earlier approach in ref. [16]
which did not yield any $MP$ features) and by a Mori calculation of the vortex density correlator (see section 5). They find three different regimes corresponding to
different ranges of the scaling variable 
\begin{equation}
      Y=\frac{1}{\omega\xi^z}
\end{equation}
$z$ being the dynmic exponent discussed in section 5.5 and $\xi$ the correlation length. The experimental data of ref. [82] fall in the intermediate regime where $MP$ is
valid. The high frequency regime corresponds to $\ll$ critical dynamics $\gg$ as it had already been exhibited in [16]. Contrary to the $\ll$ pure $MP$ $\gg$ form there
is a low frequency regime where normal Drude behaviour should be restituted. This is an important complement to $MP$, since taking the $MP$ expressions (5.5) to
(5.7) straight down to $\omega=0$, and inserting them into expression (4.45) for the electrical conductance of the array would yield superconducting low frequency
behaviour even above the transition temperature.                                                      \\

The calculation of the dynamic charge correlator $\phi_{\rho\rho}$(see section 4.2) by Mori's technique has more recently been generalized [108] by replacing the Coulomb
potential by its screened form, as it is usually done in calculations for quantum Coulomb systems [40,41]. Since screening is due to free vortices the relevant
screening length (5.13) is the $BKT$ correlation length (1.12) which is directly related to the inverse density of free vortices. This allows to impose static
critical behaviour on dynamics. The resulting friction function, determining $\phi_{\rho\rho}$ and the dielectric function has indeed a logarithmic frequency dependence, leading
to $MP$ behaviour, over a frequency interval that is increasing in length when the transition temperature is reached at which $\xi_{BKT}$ diverges. This result shows that
dynamics is $\ll$ normal $\gg$ as long as the long range Coulomb interaction is well screened by the unpaired particles and it gets more and more $MP$-like when screening
becomes less efficient (see $figure~8$). This manifests itself in a friction function that increases more and more for low frequency, making the motion more $\ll$ sluggish $\gg$. At the
transition temperature this $\ll$ critical slowing down $\gg$ is completed : the zero frequency friction function diverges.\\

    \begin{figure} [h]
\let\picnaturalsize=N
\def\picsize{14 cm}
\def\picfilename{dielectricfn.eps}
\ifx\nopictures Y\else{\ifx\epsfloaded Y\else\input epsf \fi
\let\epsfloaded=Y
\centerline{\ifx\picnaturalsize N\epsfxsize \picsize\fi \epsfbox{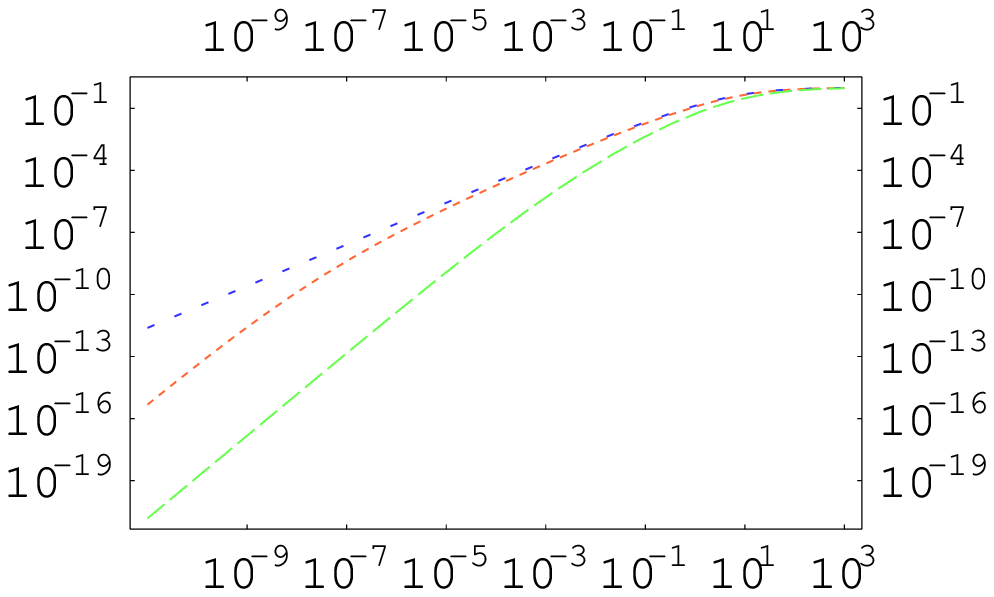}}}\fi
\label{AmplitudePhi4}
\end{figure}

    $Figure~ 8~:$ Real part of the inverse of the vortex dielectric function, as a function of frequency, calculated in the
framework of Mori's approach taking into account the motion of unbound vortices moving in a screened Coulomb
potential, for different temperatures above $T_{BKT} (T/ T_{BKT}$ = 1.02, 1.09 and 1.37 from top to bottom respectively). The frequencies are given in
units of a scale frequency $\omega_a=k_BT/\Gamma a^2$, involving the lattice constant $a$ and the bare damping $\Gamma$ (see equation
(5.10)). The slopes of the two straight lines correspond to $1/\epsilon(\omega)\sim \omega$ (Minnhagen phenomenology) and $1/\epsilon(\omega)\sim \omega^2$
(Drude behaviour).\\

While these analytic calculations treat the motion of $\ll$ free $\gg$ vortices in the force field of the Coulomb gas, the calculation of the vortex dielectric function
based on the pair relaxation picture [34] has also been refined [35] in order to study possible deviations from a simple Drude behaviour. Using a length scale
dependent static dielectric constant, obtained by numerical integration of Minnhagen's version [17] of the $BKT$ scaling relations below and above $T_{BKT}$, yields an
$\epsilon(\omega)$ that is indeed closer to $MP$ than to Drude in some intermediate frequency range.    \\

Minnhagen's group has studied $JJA$ dynamics in great detail [55,57,58,60,63] using dynamic simulations some aspects of which have been discussed in section 5.4.
These authors have used the different description (a) to (d) of $JJAs$ quoted at the beginning of section 5.4, and they have treated both unfrustrated and frustrated
arrays, in which field induced vortices are also present. They also replaced the simple cosine interaction of the phase Hamiltonian (1.2) by the form
\begin{equation}
   H=\sum_{\langle ll'\rangle}J[1-\cos^{2p^2}(\frac{\phi_{ll'}}{2})]
\end{equation}

This coupling reduces to (1.2) for $p = 1$, but for larger value of $p$ it approaches a $\ll$ potential well $\gg$ favoring the creation of vortices [109]. The dielectric
function obtained by calculating the relevant phase correlation function indeed follows closely the $MP$ prediction (see $figure~9$): contrary to the simple Drude form $Re(1/\epsilon)$ varies
linearly with $\omega,~ Im(1/\epsilon)$ has a long $\ll$ flat $\gg$ region, and the peak ratio (5.8) is close to $\pi/2$. Deviations from $MP$ for small $\omega$,
leading back to Drude, appear for $p>1$.   Frustrated systems [54,57] show similar behaviour : when the magnetic field (and with it the effective array resistance) is
increased at a fixed temperature the response crosses over from normal to anomalous. In ref. [57] it was observed that increasing the
magnetic field - ie the frustration of the array -not only leads to an obvious higher density of field induced vortices, but it also
increases the number of V-A pairs. This seems to corroborate the idea that V-A correlations are responsible for the anomlous dynamics
observed in simulations and in experiments. The screened Coulomb potential used in the Mori calculations [108] may indeed be one of the
manifestations of the presence of a sufficient number of thermal vortex excitations.   \\

\vspace{10cm}

     $Figure~ 9 ~:$ Real and imaginary parts of the inverse vortex dielectric function of an unfrustrated array using an interaction
potential with $p=2$ in equation (9.2), The open circles $(Re (1/\epsilon))$ and the filled circles $(Im(1/\epsilon))$ are the results
of dynamic simulations using $TDGL$ dynamics (method b of section 5.4) of Ref. 55. The dashed and full curves are
fits to the numerical data using the $MP$ relations (5.6).\\

The effect of (regular or random) pinning on the motion of field induced vortices of weakly frustrated JJAs has been investigated by
Monte Carlo simulation [65] of the 2D one-component lattice Coulomb gas $(OCCG)$. Anomalous MP behaviour of the dynamic dielectric
function (see subsection 5.2) is found, and it seems to be related to pinning. It is well developed when sufficiently strong random
pinning is present. In the absence of random pinning it is restricted to temperatures close to the depinning of the vortex system from
the underlying lattice structure. On the other hand, the continuum $OCCG$, which experiences no pinning at all, shows no anomalous
response. In contrast to this, the neutral 2D two-component Coulomb gas [62], which respresents the VA-system in an unfrustrated
array shows $MP$ behaviour below the $BKT$ transition. One may therefore wonder whether in the latter case vortices of one sign of charge
act as pinning centers for antivortices, and vice-versa [62].

\subsection{Flux noise}

Dynamic simulations have also been used for evaluating the flux noise spectrum either by evaluating the full integral (7.12) or by using approximate forms of the
latter, such as (7.14). At low frequency the expected white noise is found with a level that goes up when $T$ approaches $T_{BKT}$ from above.
The behaviour for higher
frequencies varies between the different authors. Some calculations [56], [110], simulating $TDGL$ dynamics for a 2D $XY$ model with a phase coupling of the type
(9.2) yield
\begin{equation}
S_{\phi}(\omega)\sim\omega^{-3/2}
\end{equation}
for some intermediate frequency range and   $S_{\phi}(\omega)\sim\omega^{-2}$  for high frequencies. The broken power of (9.3) can be understood as the sign of conventional random walk of vortices
across the boundary of the pick-up coil [110,111]. When the curves bend over from white noise to the power law (9.3) a common tangent can be drawn to the curves which
has a slope close to $1/\omega$. However, there is no extended regime showing the measured $1/\omega$ noise (see section 8.3).                \\

Other dynamic simulations [46] find an intermediate frequency range, above the white noise regime, where the spectrum follows in power law : $(1/\omega)^{\alpha}$
(see $figure~10$). The exponent $\alpha$ is $T$-dependent and differs slightly between different types of dynamics used (see section 5.4) :
\begin{equation}
\begin{array}{crl}
0.85\le\alpha\le 0.95\qquad for ~TDGL~ dynamics\cr          \\
1.17\le\alpha\le 1.27\qquad for ~RSJ~ dynamics
\end{array}
\end{equation}
which, again points to different 'critical dynamics' where different equations of motion are used.
Both cases are relatively close to the measured $1/\omega$ dependence, but - as the authors state - the $TDGL$ dynamics is closer to the measured $1/\omega$ dependence. This is
somewhat surprising, since the $\ll$usual$\gg$ picture one has of a Josephson juncion and of an array is precisely the $RSJ$ effective circuit model. It is possible that
the arrays of Shaw et al [68], consisting of large Nb islands deposited on a Cu substrate have very small resistances to ground, such that the local dissipation
term in the $TDGL$ equation dominates the bond term in $RSJ$. Other $RSJ$-simulations [112] also yield the behaviour of (9.3), both for $f=0$ and for frustrated arrays. The
same holds for the solution of the Fokker-Planck equation for pairs [43], discussed at the end of section 5.3. On the other hand a calculation [113] of the flux
noise spectrum starting from a Euclidean action for the $XY$-model with local damping also finds a cross-over from white to  $1/\omega$ noise.\\

\vspace{9cm}

          $Figure ~10~ :$ Flux noise spectra versus frequency obtained in Ref. 112 for $TDGL$ (a) and $RSJ$ (c) dynamics for different
temperatures above $T_{BKT}$. The insets of the panels (a) and (c) show the exponent $\alpha$ (see our expresssion (9.4)) as a
function of temperature. Panels (b) and (d) show a scaling plot of the noise spectrum as a function of time. The
insets present the dependence of the characteristic time t on the correlation length $x$, yielding the values $z = 2$,
respectively 0.9 of the dynamic exponent introduced in subsection 9.1.\\

The results obtained by the Mori approach to the dielectric function (section 9.2) have also been used for evaluating flux noise spectra. Preliminary
calculations [108] show that the friction function, obtained by taking into account motion of unbound particles in a screened Coulomb potential, although
reproducing nicely the MP behaviour in $\epsilon(\omega)$, does not lead to $1/\omega$ noise. These calculations are thus in good agreement with the above mentioned numerical
simulations [56], [110]. Another approach starts from the averaged pair susceptibility presented in section 5.1, integrating the single pair susceptibility $\chi$ over
a pair distribution of the form (5.2) from the lattice constant (the smallest possible pair distance) up to the $BKT$ correlation length $\xi$ and going from $\chi$ to the
density correlator $\phi_{\rho\rho}$ with the help of (4.303) yields a flux noise spectrum with a white noise level and a subsequent power law with a temperature dependent
exponent [97], similarly as in [46]. In fact, the $1/\omega$ dependence is found for $T = T_{BKT}$. The white noise level and the cross-over frequency between the latter and
the power-law regime varies like $\xi^2$. Taking the limit of zero frequency in the integral (7.10) indeed yields the following approximate statement for the white
noise level 
\begin{equation}
   S_{\phi}(\omega=0)=C\frac{\gamma(0)k_BT}{n_v(T)}
\end{equation}
where the constant is the result of the wave number integral involved in (7.10), $n_v(T)$ is the total number of vortices and $\gamma(0)$ the zero frequency limit of the
friction function. The data of reference [100] for the white noise level can be understood by limiting the integral over the pair distribution by the sample size $L$,
whenever $\xi > L$. This explains why no more critical behaviour is seen in the white noise data when the critical temperature is approached : the relevant length is
fixed by $L$ and the only major $T$-dependence left in (9.5) comes from $n_v(T)$, which is indeed thermally activated as seen in the data of ref. [100].

 \section{Driven arrays : dynamics far from equilibrium}

$JJAs$ are interesting examples of systems with non-linear equations of motion $(EM)$. The relevant $EM$ for the phases, depending on the type of dynamics, have been
found in section 2.1, whereas section 3 ends up with an equation of motion for the vortex positions. While, in sections 5 to 7, we have mainly been interested in
equilibrium dynamics and in response to weak driving fields we now turn to work dealing with the dynamic behaviour of $JJAs$ under moderate and strong driving
currents.\\

The interplay between vortex pinning due to the discrete underlying lattice structure (the $\ll$ egg cartoon potential $\gg$) and V-V interaction can be tested by changing
the applied perpendicular magnetic field, i e the frustration. In the ground state the field induced vortices tend to form a regular arrangement minimizing their
interaction as well as the pinning energy. The critical current needed to drive this vortex lattice across the sample is the higher the better the $\ll$ matching $\gg$
between array and vortex structure, which is the case for $f = p/q$ with small integers $p$ and $q$. Marconi and Dominguez have studied the depinning and melting of
driven moving vortex lattices for small frustration [114], $f=1/25$, and for $\ll$ full $\gg$ frustration [115], $f=1/2$, by solving the $RSJ$ Langevin equations of motion for the
phases in presence of the applied vector potential and a random force which maintains a fixed finite temperature. The results can be summarized in a non-equlibrium
phase diagram in the plane of temperature $T_{dep}$ and drive current $I$. For very low current it reveals the equilibrium depinning transition temperature above which the
vortex lattice floates freely on the arrays, thanks to its thermal energy. Stronger currents shift this depinning transition to lower $T$ and produce phases of
different properties above $T_{dep}$. At low $T$ the moving lattice is still $\ll$transversely pinned$\gg$ : it has anisotropic Bragg peaks, quasi-long range order and maintains
superconducting coherence (and thus a finite critical current) in the direction perpendicular to the direction of the strong drive current (one has to bear in mind
that a vortex moves in the direction perpendicular to the drive current, see section 3.1). At higher $T$ the vortex lattice is simply $\ll$ floating $\gg$ .  \\

The presence of random pinning adds new complexity. As mentioned in section 6.2, an irregular underlying lattice structure forces the vortices to $\ll$adapt$\gg$ their
equilibrium structure in order to minimize their free energy. In order to drive the vortex arrangement across the $\ll$pinning landscape$\gg$ the applied current $I$,
which acts as a transverse force on the vortices, has to exceed some threshold value $I_c$. Once the vortices are depinned, which is seen by the appearance of a
finite voltage and thus a finite resistance of the array in the current direction, their motion will, of course, be influenced by the random pinning potential.
Dominguez [116] has observed, in his numerical simulations, a dynamic transition occuring for a current $I_p (>I_c)$. For currents with a strength between $I_c$ and $I_p$
the moving $\ll$vortex liquid$\gg$ has an isotropic structure and the turbulent flow regime is plastic. Above $I_p$ there is a regime with homogeneous $\ll$laminar$\gg$ flow in
the direction of motion and with a structure with anisotropic short range order of smectic type characterized by the appearance of corresponding Bragg peaks.   \\

Various other work has been published about the possiblity of such a dynamic regime in superconducting films (for references, see for example [117]). Two dynamical
phase transitions for thin films are found theoretically in [118] : for a given strength of the exciting force the system goes again over from plastic flow, where
the vortices move in an intricate network of channels, to smectic flow. This transition is characterized by a maximum in the differential resistance (the
derivative of the averaged vortex velocity with respect to the drive current strength). At a higher driving force there is a sharp transition to a frozen
transverse solid. Here the motion of the vortices transverse to the flow direction, which consisted of jumps between different channels in the smectic regime, is
strongly reduced. More details about the temporal behaviour of the system (noise spectrum, diffusion exponents etc.) can be found in [118]. Moving anisotropic
vortex structures have also been observed experimentally [117] in small high quality single crystals of $NbSe_3$.              \\

Non-linear dynamic systems, such as our $JJAs$, can also show various phenomena related to the occurrence of chaos. Indeed, already a single, capacitively and
resistively shunted junction with an applied voltage or current corresponds to a damped driven pendulum, which is known to show very rich structures in its dynamic
phase diagram. Regimes of periodic and quasiperiodic motion alternate with chaotic regions when parameters, such as the drive force, are varied. Josephson
junctions coupled in series have a spatial variable in addition to time. They thus allow for spatio-temporal chaos. Various such systems have been studied [9]
numerically. They show coherent, ordered, partially ordered, fully turbulent and quasiperiodic phases. When time is discretized in the coupled non-linear equations
of motion for the $JJA$ the latter are of the same structure as the equations of $\ll$globally coupled maps$\gg$ [9]. Their mathematical properties (transition from
regular to chaotic motion, structure of attractors, etc.) have widely been studied. Chaotic motion in $JJAs$ should in principle be observable experimentally through
the characteristic features of the current-voltage characteristics of strongly driven arrays, but not much systematic analysis seems to have been done up to now.       \\

\section{Miscellaneous}

     Josephson junction arrays are an interesting example of articifial systems. They allow us to profit from the microscopic phenomenon of
superconductivity in order to observe challenging collective effects. This article is mainly devoted to the dynamic manifestation of the
latter. The interested reader should find elsewhere more information about the static and thermodynamic behaviour of $JJAs$ in order to
fully appreciate the rich spectrum of time dependent phenomena discussed here. For obvious reasons we had to make a choice concerning
the topics to be presented (and this choice is certainly biased by our own experience in the field !). In fact, quite a variety of other
aspects of $JJA$ dynamics have not been covered. Let us just mention a few of them:        \\

\underbar{Hall effect}     \\

In addition to the $\ll$Lorentz force$\gg$, mentioned in section 3, a vortex is also subject to a $\ll$Magnus force$\gg$ [25] when an external
charge introduces a potential between the array and the ground. This force is perpendicular to the vortex velocity. The motion resulting
from the combination of Lorentz force (showing up in equation (3.6)) and Magnus force leads to a $\ll$vortex Hall effect$\gg$ that has been
observed for classical and quantum arrays [25]. Since the Hall angle is inversely proportional to the damping constant $\Gamma$ showing up in
the vortex equation of motion, Hall measurements give information about the type of dissipation prevailing in the array. As a function
of frustration, ie of the applied magnetic field, the Hall resistance shows a rich structure [120]. Reversal of the sign of the Hall
constant, which is a sign that the relevant type of charge carriers changes, appears at several frustrations.        \\

\underbar{Ballistic vortex motion}                 \\

When the mass term in the equation of motion (3.6) dominates over the friction vortices can in principle move almost freely across the
sample. Such $\ll$ballistic$\gg$ vortex motion has been observed, both in one- and two-dimensional arrays [1,25], in a frequency window,
bounded from below by the presence of pinning forces and from above by various (non-linear) damping mechanisms setting in at high
velocities. It is interesting to note that, from a more elaborate theoretical point of view, $\ll$ideal$\gg$ ballistic motion is not really
likely to be observable [1]. Simulations on finite systems have to cope with boundary effects (a vortex being attracted by its image
when it approaches the boundary). Moreover, non-linear viscosity can lead to localized oscillations rather than propagation of the
vortex. Thus ballistic vortex motion definitely needs to be studied in more detail [1].      \\

\underbar{Shapiro steps}                  \\

In section 4.1 we have introduced a weak applied external electric field into the equations of motion and we have calculated the linear
response of the array. For an applied voltage of finite strength one can use the Josephson relation (4.19) between a voltage drop across
a junction and the corresponding time dependence of the phase difference. When the applied voltage is of the form
\begin{equation}
      V(t)=V_o+V_1\cos\omega_1t
\end{equation}
the corresponding phase difference is easily found by integrating (4.19). Substituting it into (1.6) yields an statement for the
resulting Josephson current for a single junction. Expressing the sin of a trigonometric function by a series of Bessel functions [1]
shows that one should observe a dc super current whenever
\begin{equation}
   V_o=V_n=n\frac{\hbar\omega_1}{2e}
\end{equation}
These voltage steps are called $\ll$Shapiro steps$\gg$ [121] of a single junction. Including the normal current flowing through the resistive
channel yields a current voltage characteristics showing steps at the voltage values given by (11.2). Current driven arrays are somewhat
more difficult to treat theoretically, because one really has to solve the equation of motion (2.4) for a given drive current. Shapiro
steps again occur at the voltage values of (11.2).   \\

Under $\ll$ideal$\gg$ conditions, ie in zero temperature and external field, an $M \times N$ array of such junctions should act as a single junction
with an applied current $I/M$. The total voltage drop across the array should then be $N$ times the one across each junction and $\ll$giant$\gg$
Shapiro steps should occur at voltages
\begin{equation}
\hat{V}_n=NV_n
\end{equation}
Reference 1 gives many an excellent overview over various more elaborate aspects of Shapiro steps in $JJA$, such as the occurrence of
fractional giant steps and of suharmonic steps.            \\

\underbar{Application for microwave radiation}                     \\

A Josephson junction is a voltage controlled oscillator capable of generating electromagnetic radiation [122]. The dynamics of the
supreconducting phase can be formulated in the framework of the capacitively and resistively shunted junction model presented in section
2.1. A dc bias voltage which excites the damped non-linear oscillator has to be added to the equations of motion given there.
Quantitative estimates show, however, that the linewidth of a single junction is rather large [122]. One obvious way to overcome this
drawback is now precisely to use one- or two-dimensional arrays of junctions and to have them oscillate in phase. In the 2D case treated
in this article the rf current flows, for example, along the rows of the array, whereby care has to be taken that the phases of the
current in each row are the same. The line width is then reduced, with respect to a single junction, by the number of the array sites.
Details about the realisation of such voltage controlled microwave oscillators in the frequency range between 100 $GHz$ to 1 $THz$ can be
found in reference [122].                  \\

Many of the concepts developed for arrrays can, of course, also be applied to the description of thin superconducting films which are,
in some sense, a continuum version of the discrete lattice. Thus we have mentioned results found on such films in sections 6 and 8.
Moreover, high temperature superconductors have more or less strongly anisotropic structures of rather weakly coupled lattice planes.
Thus, for many purposes the phenomenon of superconductivity can be restricted to lattice planes, so that many aspects of two-dimensional
suprerconductivity described in the present article are again applicable.               \\

{\bf{Acknowledgements}}\\

We gratefully acknowledge continuous support from the Swiss National Science Foundation and the Swiss Commission f$\acute{e}$d$\acute{e}$rale des bourses for our own
research activity in this field and
we thank D.Bormann, M.Capezzali, D.Dominguez, R.Fazio, S.Korshunov, M.Kosterlitz, P.Martinoli, P.Minnhagen, M.Mombelli, S.Shenoy,
A.Varlamov, A.Zaikin for many interesting discussions.         \\

On the other hand we express our apologies to all the many workers in the field whose results have not been adequately appreciated and
quoted in this article.      \\

{\bf{References}} \\

[1] R.S.Newrock, C.J.Lobb,U.Geigenm$\ddot{u}$ller, M.Octavio ; Solid State Physics 54, 263 \\

\hspace{0.6cm}(2000)\\

[2] P.Martinoli. C.Leemann ; Journal of Low Temperature Physics 118, 699 (2000)\\

[3]  P.W.Anderson ; in $\ll$ Lectures on the Many-Body-Problem $\gg$, Vol 2, ed. E.R.Caianiello, \\

\hspace{0.6cm}Academic Press, New York (1964)\\

[4] E. Simanek ; $\ll$ Inhomogeneous superconductors $\gg$ , Oxford University Press, 1994, \\

\hspace{0.6cm}Ch. 4 and 5\\

[5] V.L.Berezinskii ; Sov Phys JETP 32, 493 (1971)\\

[6] J.M.Kosterlitz, D.J.Thouless ; J Phys C6, 1181 (1973)\\

[7]  S.E.Korshunov, A.Vallat and H.Beck ; Phys Rev B51, 3071 (1995)\\

[8] E. Simanek ; Sol. State Comm. 31, 419 (1979)\\

[9] D. Ariosa, H.Beck, Phys Rev B45, 819 (1992)\\

[10] L.Jacobs, J.V.Jose, M.A. Novotny, A.M.Goldman, Phys Rev B38, 4562 (1988) and \\

\hspace{0.6cm}references therein\\

[11] M. Capezzali, D. Ariosa, H.,Beck; Physica B230-232, 962 (1997)\\

[12] K.H.Wagenblast, A.v Otterlo, G.Schön, G.T.Zimanyi ; Phys Rev Lett 78, 1779 (1997)\\

[13] J.M.Kosterlitz ; J.Phys C7, 1046 (1974)\\

[14] J.V. Jose et al ; Phys Rev B 16, 1217 (1977)\\

[15] A.Vallat, H.Beck ; Phys Rev B 50, 4015 (1994)\\

[16] S.R.Shenoy ; J Phys C, Solid State Physics 18, 5163 (1985)\\

[17] P.Minnhagen ; Rev Mod Phys 59, 1001 (1987)\\

[18] D.Dominguez, J.V.Jose ; Int J Mod Phys B8, 3749 (1994)\\

[19] J.C.Ciria, C.Giovanella ; J Phys Cond Mat 10, 1453 (1998)\\

[20] G.Sch$\ddot{o}$n, A.D.Zaikin ; Physics Reports 198, 237 (1990)\\

[21] V.L.Pokrovsky, M.V.Feigel'man, A.M.Tsvelick ; $\ll$ Spin waves and magnetic excita\\

\hspace{0.6cm}tions $\gg$, ed. A.S.Borovik-Romanov and S.K.Sinha ; Elsevier Science Publishers B.V. \\

\hspace{0.6cm}1988, p. 67\\

[22] H.Beck, D.Ariosa ; Solid State Communications 80, 657 (1991)\\

[23] C.J.Lobb, D.Abraham, M.Tinkham ; Phys Rev B27, 150 (1983)\\

[24] U.Eckern, A.Schmid ; Phys Rev B 39, 6441 (1981)\\

[25] R.Fazio, H.van der Zaant ; Physics reports 355, 235 (2001)\\

[26] R Fazio, U Geigenm$\ddot{u}$ller, G Sch$\ddot{o}$n ; in $\ll$ Quantum fluctuations in Mesoscopic and \\

\hspace{0.6cm}Macroscopic Systems $\gg$, eds. H.A.Caldeira et al, World scientific, Singapore, 1991, \\

\hspace{0.6cm}p. 214\\

[27] R. Fazio, G. Sch$\ddot{o}$n ; Phys Rev B 43, 5307 (1991), U Eckern, E B Sonin, Phys Rev B\\

 \hspace{0.6cm}47, 505 (1993)\\

[28] S.R.Shenoy ; J Phys C (Solid State Physics) 18, 5163 (1985)\\

[29] S.V.Panyukov, A.D.Zaikin ; Physics Letters A 156, 119 (1991)\\

[30] P.Bobbert, R.Fazio, G.Sch$\ddot{o}$n, A.D.Zaikin ; Phys Rev B 45, 2294 (1992)\\

[31] N.H.March, M.P.Tosi ; $\ll$ Atomic Dynamics in Liquids $\gg$, John Wiley and Sons, New \\

\hspace{0.6cm}York, 1976\\

[32] V.Ambegaokar, S.Teitel ; Phys Rev B 19, 1667 (1979)\\

[33] R.C$\hat{o}$t$\acute{e}$, A.Griffin ; Phys Rev B 34, 6240 (1986)\\

[34] V.Ambegaokar, B.I.Halperin, D.R.Nelson, E.D.Siggia; Phys Rev Lett 40, 783 (1978)\\

 \hspace{0.6cm}and Phys Rev B21, 1806 (1980)\\

[35] D.Bormann, H.Beck, O.Gallus, M.Capezzali ; J Phys IV France 10, Pr-447 (2000)\\

[36] H.Mori ; Prog.Theor.Phys. (Kyoto) 34, 423 (1965)\\

[37]  J.-P. Hansen, I.R. McDonald ; Theory of Simple Liquids (2nd edition), Academic \\

\hspace{0.6cm}Press Inc., San Diego, CY, 1986\\

[38] D.Forster, $\ll$ Hydrodynamic Fluctuations, Broken Symmetry and Correlation \\

\hspace{0.6cm}Functions $\gg$, Benjamin/Cummings,s London, 1975\\

[39] N.G. van Kampen ;  $\ll$ Stochastic Processes in Physics and Chemistry $\gg$, North-\\

\hspace{0.6cm}Holland, Amsterdam, 1981\\

[40] R.Zimmermann, H.Stolz ; Phys Stat Sol (b) 131, 151 (1985)\\

[41] H.E.DeWitt, M.Schlanges, A.Y.Sakakura, W.D.Kraeft ; Phys Lett A197, 326 (1995)\\

[42]  M.Capezzali, M.Mombelli, P.B$\acute{e}$ran, H.Beck ; in $\ll$Macroscopic Quantum Phenomena\\

\hspace{0.6cm} and Coherence in Superconducting Networks $\gg$,ed. C.Giovannella and M.Tinkham, \\

\hspace{0.6cm}World Scientific, Sikngapore, 1995, p. 270\\

[43] C.Timm ; Phys Rev B 55, 3241 (1997)\\

[44] B.J. Kim, P Minnhagen, P. Olson ; cond-mat/9806231, Phys Rev B 59, 11'506 (1999)\\

[45] L.M. Jensen, B.J. Kim, P. Minnhagen ; Europhys Lett 49, 644 (2000) ; Phys Rev B\\

\hspace{0.6cm}61, 15'412 (2000)\\

[46] P.H.E.Tiesinga, T.J.Hagenaars, J.E.van Himbergen, J.V.Jose ; Phys Rev Lett 78, 519 \\

\hspace{0.6cm}(1997)\\

[47] I-J.Hwang, S.Ryu, D.Stroud ; cond-mat/9704108\\

[48] B.J.Kim, P.Minnhagen ; cond-mat/9902148 ; Phys Rev B 60, 6834 (1999)\\

[49] P.Minnhagen, O.Westman, A.Jonsson, P.Olsson ; Phys Rev Lett 74, 3672 (1995)\\

[50] L.M.Jensen, B.J.Kim, P. Minnhagen ; cond-mat/0003447 ; Phys Rev B 61, 15412 \\

\hspace{0.6cm}(2000)\\

[51] M.V.Simkin, J.M.Kosterlitz ; Phys Rev B 55, 11'646 (1997)\\

[52] I-J Hwang, D.Stroud ; Phys Rev B 57, 6036 (1998)\\

[53] B.J.Kim, P.Minnhagen, P.Olsson ; cond-mat/9806231 ; Phys Rev B 59, 11506 (1999)\\

[54] B.J.Kim, P.Minnhagen ; Phys Rev B 60, R15'043 (1999)\\

[55] A. Jonsson, P. Minnhagen ; Phys Rev B 55, 9035 (1997)\\

[56] J.Houlrik, A.Jonsson, P.Minnhagen ; Phys Rev B 50, 3953 (1994)\\

[57] A.Jonsson, P.Minnhagen ; Physica C 277, 161 (1997)\\

[58] A.Jonsson, P.Minnhagen ; Phys Rev Lett 73, 3576 (1994)\\

[59] P.Minnhagen, B.J.Kim, H.Weber ; Phys Rev Lett 87, 037002 (2001)\\

[60] P. Minnhagen, O.Westman ; Physica C 220, 347 (1994)\\

[61] K.Holmlund, P.Minnhagen ; Physica C 292, 255 (1997)\\

[62] K.Holmlund, P.Minnhagen ; Phys Rev B 54, 523 (1996)\\

[63] P.Minnhagen ; Physica B 222, 309 (1996)\\

[64] H.Weber, M.Wallin, H.J.Jensen ; Phys Rev B 53, 8566 (1996)\\

[65] B.J.Kim, P.Minnhagen ; cond-mat/0011297 (Nov 2000)\\

[66] P.Olsson ; Phys Rev B 46, 14'598 (1992), Phys Rev B 52, 4511 (1995), Phys Rev B, \\

\hspace{0.6cm}4526 (1995)\\

[67] P.C.Hohenberg, B.I.Halperin ; Rev Mod Phys 49, 435 (1977)\\

[68] T.J.Shaw, M.J.Ferrari,L.L.Sohn, D.H.Lee, M.Tinkham, J.Clarke ; Phs Rev Lett 76,\\

\hspace{0.6cm} 2551 (1996)\\

[69] D.S.Fisher, M.P.A.Fisher, D.A.Huse ; Phys Rev B 43, 130 (1991) ; A.Dorsey, Phys \\

\hspace{0.6cm}Rev B 42, 7575 (1991)\\

[70] D.Bormann ; Phys Rev Lett 78, 4324 (1997)\\

[71] S.M.Ammirata, M.Friesen, St.W.Pierson, LeRoy A.Gorham, J.C.Hunnicutt, M.L.Trawick, \\

\hspace{0.6cm}C.D.Keener ; Physica C313, 225 (1999)\\

[72] T.Nakayama, K.Yakubo, R.L.Orbach ; Rev Mod Phys 66, 381 (1994)\\

[73] M.Mombelli, H.Beck ; Phys Rev B57, 14'397 (1998)\\

[74] A.L.Efros, B.I.Shklovskii ; Phys Status Solidi B76, 475 (1976)\\

[75]  A.M.Dykhne, Zh.Eksp.Teor.Fiz. 59, 110 (1970) [Sov.Phys.JETP 32, 63 (1971)], \\

\hspace{0.6cm}J.P.Straley, Phys Rev B 15, 5733 (1977)\\

[76] M.P.A.Fisher, Phys Rev Lett 62, 1415 (1989)\\

[77] M.V.Feigel'man, V.B.Geshkenbein, A.I.Larkin, V.M.Vinokur; Phys Rev Lett 63, 2303 \\

\hspace{0.6cm}(1989)\\

[78] M.Calame, S.E.Korshunov, Ch.Leemann, P.Martinoli ; Phys Rev Lett  86, 3630 (2001)\\

[79] S.T.Herbert, Y.Jun, R.S.Newrock, C.J.Lobb, K.Ravindran, H.K.Shin, D.B.Mast, \\

\hspace{0.6cm}S.Elhamri ; Phys Rev B 57, 1154 (1998)\\

[80] A.F.Hebard, A.T.Fiory ; Phys Rev Lett 44, 291 (1980)\\

[81] P.Martinoli, Ph.Lerch, Ch.Leemann, H.Beck ; Japanese Journal of Applied Physics \\

\hspace{0.6cm}26-3, 1999 (1987)\\

[82] R.Th$\acute{e}$ron, J.B.Simond, C.Leemann, H.Beck, P.Martinoli, P.Minnhagen ; Phys Rev \\

\hspace{0.6cm}Lett 71, 1246 (1993)\\

[83] B.Jeanneret, J.L.Gavilano, G.A.Racine, Ch.Leemann, P.Martinoli ; Appl Phys Lett \\

\hspace{0.6cm}55, 2336 (1989)\\

[84] Ch.Leemann, Ph.Lerch, G.A.Racine, P.Martinoli ; Phys Rev Lett 56, 1291 (1986)\\

[85] B.Jeanneret, Ph.Fluckiger, Ch.Leemann, P. Martinoli ; Jpn J Appl Phys 26-3, 1417 \\

\hspace{0.6cm}(1987)\\

[86] S.Korshunov ; unpublished\\

[87] H.Weber, M. Wallin, H.J.Jensen ; Phys Rev B53, 8566 (1996)\\

[88] A.M.Kadin, K.Epstein, A.M.Goldman, Phys Rev B27, 6691 (1983),  P.Minnhagen, \\

\hspace{0.6cm}Phys Rev B 28, 2463 (1983)\\

[89] I.G.Gorlova, Yu.I.Latyshev, Pis'ma Zh Eksp Teor Fiz 51, 197 [(1990) JETP Lett 51, \\

\hspace{0.6cm}224 (1990)]\\

[90] A.K.Pradhan, S.J.Hazell, J.W.Hodby, C.Chen, Y.Hu, B.M.Wanklyn ; Phys Rev B47, \\

\hspace{0.6cm}11'374 (1993)\\

[91]  J.E.Mooij ; in $\ll$ Percolation, Localization and Superconductivity $\gg$, ed. A.M.Goldman \\

\hspace{0.6cm}and S.A.Wolf  (Plenum Press, New York, 1984) NATO ASI SERIES Vol. 8109, p. 325\\

[92] P.G. DeGennes ; Rev Mod Phys 36, 225 (1964)\\

[93] P.Minnhagen ; Phys Rev B 32, 7548 (1985)\\

[94] J.Holzer, R.S.Newrock, C.J.Lobb, T.Aouaroun, S.T.Herbert ; preprint (2001)\\

[95] A.L.Eichenberger, J.Affolter, M.Willemin, M.Mombelli, H.Beck, P. Martinoli, \\

\hspace{0.6cm}S.E.Korshunov; Phys Rev Lett 77, 3905 (1996)\\

[96] Fractal and Disordered Systems, ed. A.Bunde and S.Havlin (Springer-Verlag, Berlin, \\

\hspace{0.6cm}1991)\\

[97] Md.Ashrafuzzaman, H.Beck ; unpublished\\

[98] J. Affolter ; Ph-Thesis, University of Neuch$\acute{a}$tel (2001)\\

[99] S.E.Korshunov ; cond-mat/0007387, and references therein\\

[100] S.Candia, Ch.Leemann, S.Mouaziz, P.Martinoli ; cond-mat/0111212, to appear in \\

\hspace{0.6cm}Physica C\\

[101] O.Festin, P.Svendlindh, B.J.Kim, P.Minnhagen, R.Chakalov, Z.Ivanov ; Phys Rev \\

\hspace{0.6cm}Lett 83, 5567 (1999)\\

[102] C.T.Rogers, K.E.Myers, J.N.Eckstein, I.Bozovic ; Phys Rev Lett 69, 160 (1992)\\

[103] H.Beck ; Phys Rev B 49, 6153 (1994)\\

[104] S.E.Korshunov ; Phys Rev B 50, 13616 (1994)\\

[105] U.Eckern, A.Schmid ; Phys Rev B 39, 6441 (1989)\\

[106] U.Geigenm$\ddot{u}$ller, C.J.Lobb, C.B.Whan ; Phys Rev B 47, 348 (1993)\\

[107] M.Capezzali, H.Beck, S.R.Shenoy ; Phys Rev Lett 78, 523 (1997)\\

[108] M.Capezzali, H.Beck ; unpublished\\

[109] E.Domani, M.Schick, R.Swendson ; Phys Rev Lett 52, 1535 (1984)\\

[110] B.J.Kim, P.Minnhagen ; Phys Rev B 60, 6834 (1999)\\

[111] R.F.Voss, J.Clark ; Phys Rev B 13, 556 (1976)\\

[112] I.J.Hwang, D.Stroud ; Phys Rev B 57, 6036 (1998)\\

[113] K.H.Wagenblast, R.Fazio ; cond-mat/9611177 ; JETP Lett (USA)68, 312 (1998), \\

\hspace{0.6cm}english translation of Pis'ma Zh Eksp Teor Fiz (Russia) 68, 291 (1998)\\

[114] V. I. Marconi and D. Dominguez, Phys. Rev. B 63, 174509 (2001). This article also \\

hspace{0.6cm}contains a rather exhaustive list of other references\\

[115] V. I. Marconi and D.Dominguez, Phys. Rev. Lett. 87, 017004 (2001).\\

[116] D.Dominguez ; Phys Rev Lett 82, 181 (1999)\\

[117] F.Pardo, F.de la Cruz, P.L.Gammel, E.Bucher, D.J.Bishop; Nature (London) 396, \\

\hspace{0.6cm}348 (1998)\\

[118] A.B.Kolton, D.Dominguez, N.Gronbech-Jensen ; Phys Rev Lett 83, 3061 (1999)\\

[119] D.Dominguez, H.A.Cardeira ; Phys Lett A200, 43 (1995) ; Phys Rev B 52, 513 \\

\hspace{0.6cm}(1995) ; and other references quoted therein\\

[120] C.D.Chen, P.Delsing, D.B.Haviland, T.Claeson ; in $\ll$ Macroscopic Quantum \\

\hspace{0.6cm}Phenomena and Coherence in Superconducting Networks $\gg$, World Scientific, \\

\hspace{0.6cm}Singapore (1995), p. 121\\

[121] S.Shapiro ; Phys Rev Lett 11, 80 (1963)\\

[122] J.Mygind, N.F.Pedersen ; in $\ll$ Macroscopic Quantum Phenomena and Coherence \\

\hspace{0.6cm}in Superconducting Networks $\gg$, World Scientific, Singapore (1995), p. 339.This \\

\hspace{0.6cm}book also contains other references dealing with microwave radiation from JJAs.\\

\end{document}